\documentclass[fleqn,10pt]{wlscirep}
\usepackage[utf8]{inputenc}
\usepackage[T1]{fontenc}
\title{Reconstructing dynamics of complex systems from noisy time series with hidden
variables}

\author[1]{Zishuo Yan}
\author[2]{Lili Gui}
\author[2]{Kun Xu}
\author[1,2,*]{Yueheng Lan}
\affil[1]{School of Science, Beijing University of Posts and Telecommunications, Beijing,100876, China}
\affil[2]{State Key Lab of Information Photonics and Optical Communications,
Beijing University of Posts and Telecommunications, Beijing,100876, China }

\affil[*]{lanyh@bupt.edu.cn}

\begin{abstract}
Reconstructing the equation of motion and thus the network topology of a system from time series is a very important problem. Although many powerful methods have been developed, it remains a great challenge to deal with systems in high dimensions with partial knowledge of the states. In this paper, we propose a new framework based on a well-designed cost functional, the minimization of which transforms the determination of both the unknown parameters and the unknown state evolution into parameter learning. This method can be conveniently used to reconstruct structures and dynamics of complex networks, even in the presence of noisy disturbances or for intricate parameter dependence. As a demonstration, we successfully apply it to the reconstruction of different dynamics on complex networks such as coupled Lorenz oscillators, neuronal networks, phase oscillators and gene regulation, from only a partial measurement of the node behavior. The simplicity and efficiency of the new framework makes it a powerful alternative to recover system dynamics even in high dimensions, which expects diverse applications in real-world reconstruction.
\end{abstract}
\begin{document}

\flushbottom
\maketitle
%\thispagestyle{empty}

%\noindent Please note: Abbreviations should be introduced at the first mention in the main text – no abbreviations lists. Suggested structure of main text (not enforced) is provided below.

\section*{Introduction}
Complex interacting systems are ubiquitous and essential for the modern world which include both the social ones: financial systems, power systems, transportation systems, the internet, and the natural ones: neural networks, gene regulation systems and the earth system~\cite{1Sch1992Nonlinear,2Winkel1995Application,3pasten2018time,4Gouveia2000Time,5caldarelli2013reconstructing,62009The,7chen2014the,81998Genome,92003Inferring,102006algorithm,2009Inferring,2012Model}. Diverse and intriguing behaviour is seen arising, out of the complexity of dynamics of individual agents constituting the network, or more often, of the interactions between them. Although more and more data are now measured in these systems, it remains a great challenge to reconstruct network structures and dynamics because a direct measurement of them is rarely possible~\cite{B2015Systems,2006Complex,2004Forecast,2009Complex,Mcharakspar2017Dynamic,2003Newman}, which nevertheless are critical for an effective description, analysis, prediction and control of system behaviour. 
\par
System reconstruction is an inverse problem that has been studied extensively in different contexts at various levels~\cite{33PhysRevLett,34shandilya2011inferring,352008Reconstructing,2016Discovering,2017Model}. With the time-delay embedding technique~\cite{Lorenz1991DimensionOW}, an equivalent phase space may be constructed and many dynamical features could be computed but interaction rules between different coordinates need further exploration. Another set of convenient tools are based on information theoretics~\cite{112007Partial,122011Momentary,Sch2000,522008Kernel}, which quickly deduce the network topology from time series but the exact strengths of connections are hard to retrieve.  Moreover, with these techniques, a computation of direct interaction links could be very cumbersome in complex networks~\cite{2012Model}.  To obtain equations of motion, model-based methods are used to determine unknown parameters in the pre-assumed dynamical models, such as differential equations or discrete maps. However, many of these methods only apply to low-dimensional systems~\cite{1992Fitting,2017Brunton}.  To conveniently reconstruct networked systems, methods based on cybernetics were designed~\cite{13yu2006estimating,142007Topology,152008Synchronization,162009Structure,17Parlitz1996Estimating,182007Estimating,43Zoran2011Network}, so that the auxiliary response network synchronizes with the studied network (driving network) via the measured time series, if the network parameters are correctly estimated. However, when applying these methods, we should know the node dynamical equations and measured the states of all nodes with no interference, which may not be possible in practice. 
\par
Almost invariably, noise is an unavoidable factor in almost all practical reconstruction, which constitutes a major difficulty on various occasions.  Distinct methods have been developed to remove or utilize noise in a time series to recover the interaction patterns based on noise-induced dynamics or correlations~\cite{19ren2010noise,20Emily2014Erratum,21ching2017reconstructing,22zhang2015solving,232017Reconstruction}, in which either a small noise limit is assumed or priori smoothing of the data should be performed. To utilize all the data to suppress fluctuations, Wang~\cite{2021Reconstruction} proposed a globally invariant local fitting method, which is capable of processing data with moderate or strong noise. For high-dimensional systems, if the structure of the complex network is known, Gao~\cite{2022Autonomousinference} developed a two-stage approach (first stage, global regression; second stage, local fine-tuning) for autonomous inference of node and interaction dynamics of complex systems, which, however, needs knowledge of the network topology and the state of all nodes. 
\par
Lack of measurement of all variables is surely another major problem for dynamics reconstruction, which requires the recovery of not only unknown system parameters but also the time course of unmeasured variables - the so-called hidden variables.
In 1990, Breeden {\em et al.}~\cite{24201990Reconstructing} discussed how to reconstruct system equations in the presence of hidden variables. Their method is based on the flow algorithm developed by Cremers {\em et al.}~\cite{251987Construction}, which is similar to that of Crutchfield
and McNamara\cite{26James1987Equations}. It can be applied to systems with one or more such hidden variables, provided that the form of the reconstruction function is known. In a different trial, Gouesbet~\cite{27gouesbet1991reconstruction,28gouesbet1992reconstruction}
introduced the concept of standard system to numerically evaluate a set of unknown constants related to the target vector field but high-order derivatives are needed which may be hard to compute in the presence of noise.  Wang {\em et al.}~\cite{2011Network,292012Detecting,302014Uncovering,31Ri2016Data,322014Reconstructing} developed a completely data-driven method by using compressed sensing, which effectively infers the existence and location of hidden nodes in a network based on anomalous prediction errors. However, the determination of the coupling strength requires further investigation. With an information theoretic approach, Wu~\cite{36Xiaoqun2012Inferring} used the piecewise approximation to carry out a partial Granger causality test (Guo \cite{37GUO200879}) to detect the hidden variables with minor impact. The method of high-order cross correlation (HOCC)~\cite{38chen2017reconstruction,39zhang2017network,41Wangsh} is proposed as a general technique to reconstruct networked systems driven by noise and with hidden variables. However, the requirement of high-order derivatives could easily amplify noise, which fails the reconstruction if the noise is not small.  Similar methods were developed and explained by Ching and Tam~\cite{2013Extracting,21ching2017reconstructing,442018Effects} based on observable covariances.
\par
In all, the presence of hidden variables brings considerable challenges to the network dynamics reconstruction, especially when the system dimension is high and the dynamics is contaminated with noise. Here we propose a new framework to do the reconstruction based on a cost functional, which infers both the unknown parameters and unmeasured time series of hidden variables simultaneously. A gradient descent algorithm is employed and leads to a set of ordinary differential equations (ODEs). The measured time series are used as inputs which drive these ODEs such that the cost functional approaches to zero. Thus it is possible to recover the parameters and trajectories in real time with online input. Even in the presence of noise and in high dimensions, the proposed reconstruction scheme is still valid. As expected, the more the measured data, the less the derived equations. Because of the evolution character of the method, different from most of the current existing methods, the unknown parameters could enter the equations in a nonlinear way. Successful application of the method to typical examples of network models demonstrated its efficiency and great potential in dynamics determination in complex systems. 
\par
The rest of the paper is organized as follows. The main ideas of our approach are presented in detail in Section II. In Section III, we first apply the method to the well-known Lorenz system driven by white noise, and check the impact of noise intensity and parameter selection. Then, its validity is further demonstrated in application to networks with different local dynamics: the Lorenz oscillators or the FitzHugh-Nagumo(FHN) neural network models. Finally, a discussion of hidden nodes and applicability of the method is made with an example model of coupled Kuramoto oscillators. In the final section, we conclude this paper and point out possible future directions.
\section*{Results}
\subsection*{Theory}
\label{theory}
We consider a system consisting of $N$ nodes, where the dynamics of each node is determined by a set of differential equations with interactions from other nodes that  form a network, given by:
\begin{align}
 \label{eq:f1}\dot{\mathbf{X}_{i}} & = \bold{F}_i(\bold{X},\bold{p})=\bold{f}_{i}(\bold{X}_{i},\bold{B}_{i})+\sum_{j=1}^{N}A_{ij}\bold{g}_{ij}(\bold{X}_{i},\bold{X}_{j})
\end{align}
where $i,j \in {1,2,3\cdots,N}$ and $\mathbf{X}_{i}=\left[X_{i}^{(1)}, X_{i}^{(2)}, \ldots, X_{i}^{(D)}\right] \in \mathbb{R}^{\mathrm{D}}$ is the state vector of node $i$. The vector $\mathbf{F}_i$ dictates the motion of the $i$-th node with $\mathbf{f}_{i}, \mathbf{g}_{ij} $being smooth functions. The parameters $\mathbf{p}$ include the local ones $\mathbf{B}_i$ and the coupling strength $A_{ij}$. The functional form $\mathbf{f}_{i}$ and $\mathbf{g}_{ij}$ are assumed known but the parameters $\mathbf{p}$ have to be determined from the given time series.
\begin{figure}[htbp]
	\centering
	\includegraphics[height=7cm,width=0.7\textwidth]{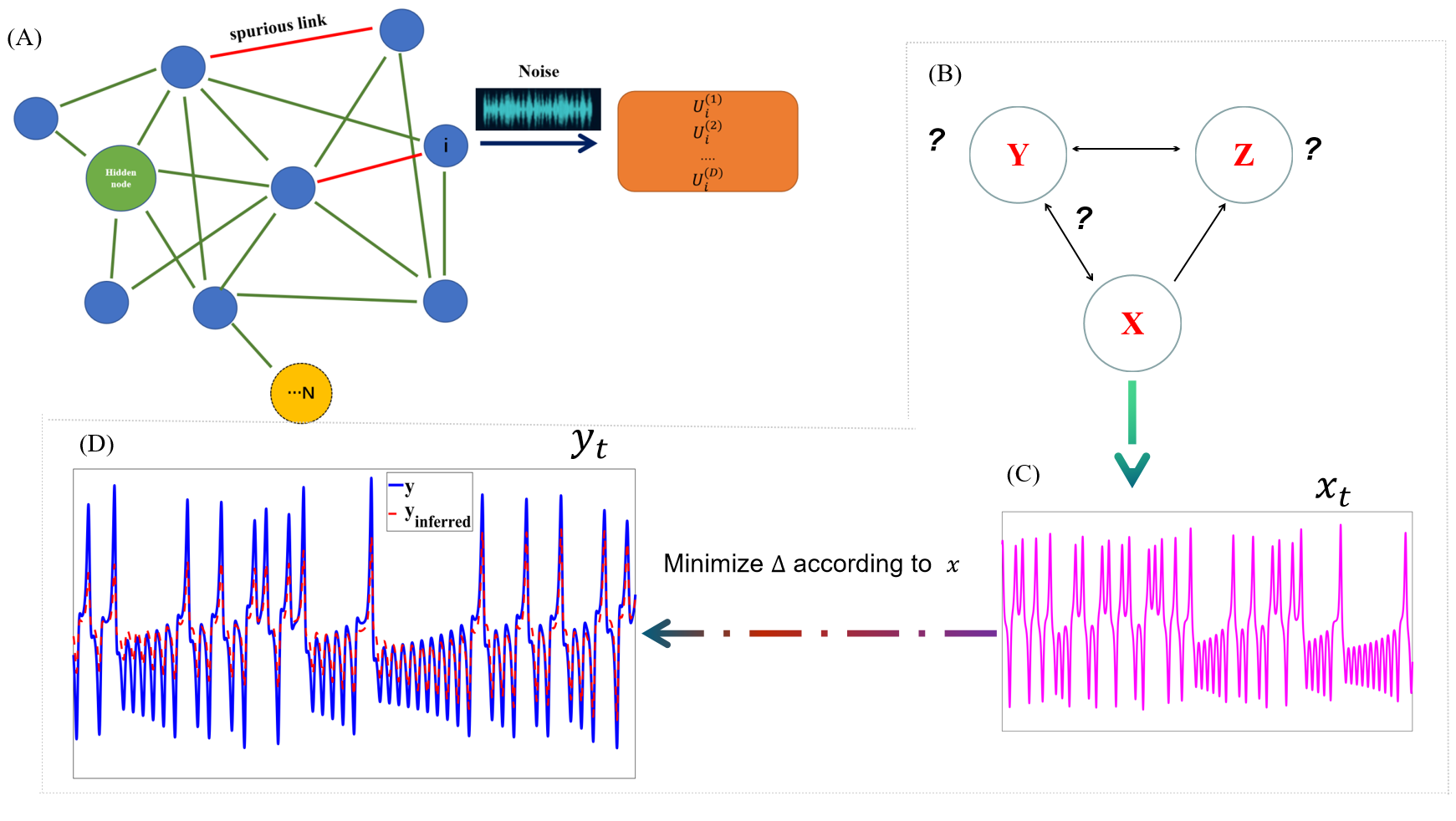}
	\caption{Illustration of a complex network. (A) The number of nodes of the network is $N$, the $i$-th of which has a generic dimension $D_i$. In practice, not all nodes can be measured, so our aim is to infer all the unknown parameters including the coupling strength from noisy time series in the presence of hidden variables (nodes) or even spurious links (the green node). ${U}_{i}$ indicates the measured noisy data. (B) A simple example of graph (A), where the time series of Y, Z and the coupling strength are unknown, and the time series of X (C)is known. Based on the known time series X, the real time series of Y can be slowly converged from any initial value by minimizing the $\Delta$ in Eq.~(\ref{eq:f2b}).}
\label{fig:fnet}
\end{figure}
\par
In the following, we present a method for reconstructing the network connection and equation parameters with only a partial observation of the system state. Those measurable variables constitute an $n$-dimensional subspace, denoted by $\mathbf{x}$ and the complementary space is thus $N \times D-n$-dimensional marked with variables $\mathbf{y}$. Of course, we may write $\mathbf{X}^t=(\mathbf{x}^t,\mathbf{y}^t)$ if $\mathbf{x},\mathbf{y}$ are properly arranged. Their dynamics is governed by Eq.~(\ref{eq:f1}), which with a slight abuse of notations could be written in a more convenient form
\begin{align}
 \label{eq:f3}\dot{\mathbf{x}}=\mathbf{F}_{o}(\mathbf{x},\mathbf{y},\mathbf{p})\,,
\dot{\mathbf{y}}=\mathbf{F}_{h}(\mathbf{x},\mathbf{y},\mathbf{p}) 
\,,
\end{align}
where the subscripts $o=(1,2,\cdots,n)^t$ and $h=(n+1,n+2,\cdots,N*D-n)^t$ denote different dimensions of the phase space.

To reconstruct the system, an error function is proposed as an integrated squared difference between the left- and the right- handside of Eq.~(\ref{eq:f3}) 
\begin{align}
 \label{eq:f2}\Delta= \int_{0}^{t}(\dot{\mathbf{x}}-\mathbf{F}_{o}(\mathbf{x},\mathbf{y},\mathbf{p}))^{2}+\beta(\dot{\mathbf{y}}-\mathbf{F}_{h}(\mathbf{x},\mathbf{y},\mathbf{p}))^2 d s
\end{align}
in a time interval $[0\,,t]$, where $\beta$ is an adjustable hyper-parameter.  With correct values of the parameters and orbit points, obviously $\Delta$ takes its minimum value $0$, indicating the existence of an absolute minimum.  While the velocity $\dot{\mathbf{x}}$ could be computed from the known time series of $\mathbf{x}(t)$~\cite{2021Reconstruction}, the state variables $\mathbf{y}$ as well as their time variations $\dot{\mathbf{y}}$ have to be deduced.
The strategy is to systematically modify the parameters $\mathbf{p}$ and the initial conditions of $\mathbf{y}$ so as to minimize the cost function Eq.~(\ref{eq:f2}), treating $\mathbf{x}$ and $\dot{\mathbf{x}}$ as known.

The integration time $t$ should be long enough to exhibit characteristics of the orbit and to suppress small noise in the data. Nevertheless, if $t$ is really large, due to the exponential amplification of deviations in a chaotic system, the error function $\Delta$ will become extremely sensitive to the parameters as well as to the initial conditions. To balance the computation, an exponentially decaying factor $e^{-\alpha(t-s)}$ is supplied to the integrand. With this modification, the error function essentially depends only on a time interval ending at $t$ and with a length characterized by the hyper-parameter $\alpha$. With this decaying factor, the memory of initial conditions becomes vague with increasing $t$ and we have to employ an alternative way to push the state point to the correct trajectory. For this, we modify the equation of motion of $\mathbf{y}$ in the following manner
\begin{align}
 \label{eq:f3b}
\dot{\mathbf{y}}=\mathbf{F}_{h}(\mathbf{x},\mathbf{y},\mathbf{p})+\mathbf{e} 
\,,
\end{align}
where $\mathbf{e}$ are artificial parameters to gradually correct errors due to the wrong initial conditions, which has the same dimension as $\mathbf{y}$.
Later on, as will be seen, the parameters $\mathbf{e}$ are also evolving slowly, sending $\mathbf{y}$ to the right track. Toward the end of the course, $\mathbf{e}$ gradually shrinks to $0$, recovering the original equations of motion Eq.~(\ref{eq:f3}). 

With all the above consideration, the error functions could be modified to
\begin{align}
 \label{eq:f2b}\Delta= \int_{0}^{t}e^{-\alpha(t-s)}[(\dot{\mathbf{x}}-\mathbf{F}_{o}(\mathbf{x},\mathbf{y},\mathbf{p}))^{2}+\beta\mathbf{e}^2] d s
 \,.
\end{align}
Thus, the problem converts to parameter learning which minimizes the function Eq.~(\ref{eq:f2b}). Nevertheless, it is not as simple as it seems. The state variable $\mathbf{y}$ is in fact an implicit function of $\mathbf{p}$ and $\mathbf{e}$ since it is computed from the evolution equation
~(\ref{eq:f3b}). Thus, we have
\begin{footnotesize}  
\begin{equation}
\begin{aligned}
\label{eq:f4}
&\frac{\partial \Delta}{\partial \mathbf{p}}=\int_{0}^{t} e^{-\alpha(t-s)}\sum_{i=1}^{n}-2 \left(\dot{x}_{i}-F_{i}(\mathbf{x}, \mathbf{y}, \mathbf{p})\right)\left(\frac{\partial F_{i}(\mathbf{x}, \mathbf{y}, \mathbf{p})}{\partial \mathbf{y}}  \frac{\partial \mathbf{y}}{\partial \mathbf{p}}+\frac{\partial F_{i}(\mathbf{x}, \mathbf{y}, \mathbf{p})}{\partial \mathbf{p}}\right) d s \equiv \mathbf{H}(t)
\\
&\frac{\partial \Delta}{\partial \mathbf{e}}=\int_{0}^{t} e^{-\alpha(t-s)}\left(\sum_{i=1}^{n}-2 \left(\dot{x}_{i}-F_{i}(\mathbf{x}, \mathbf{y}, \mathbf{p})\right)\left(\frac{\partial F_{i}(\mathbf{x}, \mathbf{y}, \mathbf{p})}{\partial \mathbf{y}}  \frac{\partial \mathbf{y}}{\partial \mathbf{e}}\right) +2  \beta e\right) d s \equiv \mathbf{E}(t)
\,.
\end{aligned}
\end{equation}
\end{footnotesize} 

From Eq.~(\ref{eq:f3b}), we may get the evolution of the parameter dependence 
\begin{footnotesize}  
\begin{equation}
\begin{aligned}
\label{eq:f3c}
&\frac{d}{dt}(\frac{\partial \mathbf{y}}{\partial \mathbf{p}})=\frac{\partial \mathbf{F}_{\mathbf{h}}}{\partial \mathbf{y}}\frac{\partial \mathbf{y}}{\partial \mathbf{p}}+\frac{\partial \mathbf{F}_{\mathbf{h}}}{\partial \mathbf{p}}-\alpha\frac{\partial \mathbf{y}}{\partial \mathbf{p}}
\\
&\frac{d}{dt}(\frac{\partial \mathbf{y}}{\partial \mathbf{e}})=\frac{\partial \mathbf{F}_{\mathbf{h}}}{\partial \mathbf{y}}\frac{\partial \mathbf{y}}{\partial \mathbf{e}}+1-\alpha\frac{\partial \mathbf{y}}{\partial \mathbf{e}}
\,.
\end{aligned}
\end{equation}
\end{footnotesize} 
In the above two equations, the last terms that contain $\alpha$ are supplied to damp the parameter dependence of earlier times. With the increase of integration time, this dependence could become extremely sensitive in a chaotic motion if $\alpha=0$. Therefore, those two terms will keep a finite memory as done in Eq.~(\ref{eq:f2b}). 

The evolution of parameters can be chosen along the gradient direction, thus 
\begin{equation}
\label{eq:f8}
\frac{d \mathbf{p}}{d t}=-2 \gamma \mathbf{H} \,,\frac{d \mathbf{e}}{d t}=-2 \gamma \mathbf{E}
\end{equation}
where $\gamma$ is the learning rate and the variables $\mathbf{H}\,,\mathbf{E}$ are defined in Eq.~(\ref{eq:f3b}) and satisfy the equations
\begin{equation}
\label{eq:f9}
\begin{aligned}
\frac{d \mathbf{H}}{d t}=-\alpha \mathbf{H}+\mathbf{H}_1\,,\frac{d \mathbf{E}}{d t}=-\alpha \mathbf{E}+\mathbf{E}_1
\end{aligned}
\,,
\end{equation}
where $\mathbf{H}_1\,,\mathbf{E}_1$ is the integrand apart from the decay factor in the expression $\mathbf{H}\,,\mathbf{E}$ displayed in Eq.~(\ref{eq:f4}). 

With the above set of evolution equations for $\mathbf{y}\,,\frac{\partial \mathbf{y}}{\partial \mathbf{p}}\,,\frac{\partial \mathbf{y}}{\partial \mathbf{e}}\,,\mathbf{p}\,,\mathbf{e}\,,\mathbf{H}\,,\mathbf{E}$ and the given time series of $\mathbf{x}$, we may drive the unknown variables to recover both the parameters and the orbit. The initial conditions of these equations are given somewhat arbitrarily although prior information could be used to speed up the convergence. Assuming the number of unknown parameters $\mathbf{p}$ is $n_p$, then the total number of parameters is $n_c=n_p+n_e=n_p+N*D-n$ where $n_e=N*D-n$ is the number of the auxiliary parameters $\mathbf{e}$. Therefore, the total number of equations to solve is $n_t=(N*D-n)(1+n_c)+2n_c$, since we care about each hidden variable as well as their derivatives with respect to all the parameters and the evolution of each parameter needs two equations(check Eq.~(\ref{eq:f8})and Eq.~(\ref{eq:f9})).  It is easy to see that the number of equations is quadratic in the number of hidden variables $\mathbf{y}$ but linear in unknown parameters $\mathbf{p}$. Hence, the more information we have, the less effort is needed to solve these equations, being consistent with the common intuition. Thus, the complexity increases quadratically with the number of hidden variables, which makes this type of problems very hard.

\subsection*{Numerical simulations}
In the following, several well-known examples are used to illustrate application of the above formulation for recovery of system dynamics and parameters. The absolute error of an estimated parameter is calculated as $\lvert parm-parm'\rvert$, where $parm$ is the inferred result and $parm'$ is the true value. First, as usual we try the Lorenz system and check how the performance of the new algorithm depends on various hyper-parameters. 
\subsubsection*{Lorenz system}
\begin{figure}[!htbp]
	\centering
	\includegraphics[height=7cm,width=0.7\textwidth]{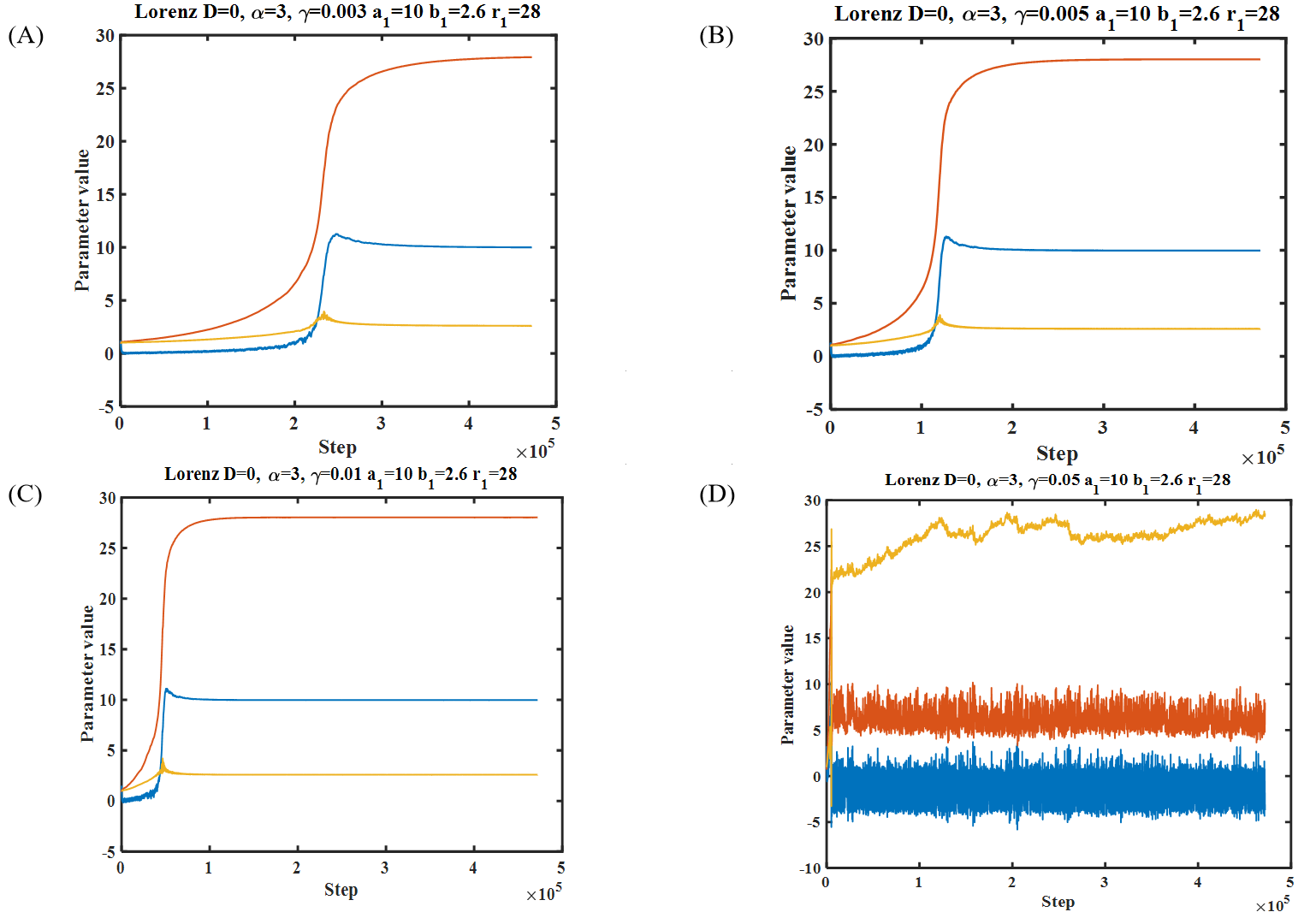}
	\caption{Reconstruction with different $\gamma$ at $t_ {unit} = 0.01,\alpha=3$, and $D = 0$. (A) $\gamma=0.003$. (B) $\gamma=0.005$. (C) $\gamma=0.01$. (D) $\gamma=0.05$.}
	\label{fig:f1}
\end{figure}

\begin{figure}[htbp]
	\centering
	\includegraphics[height=7cm,width=0.7\textwidth]{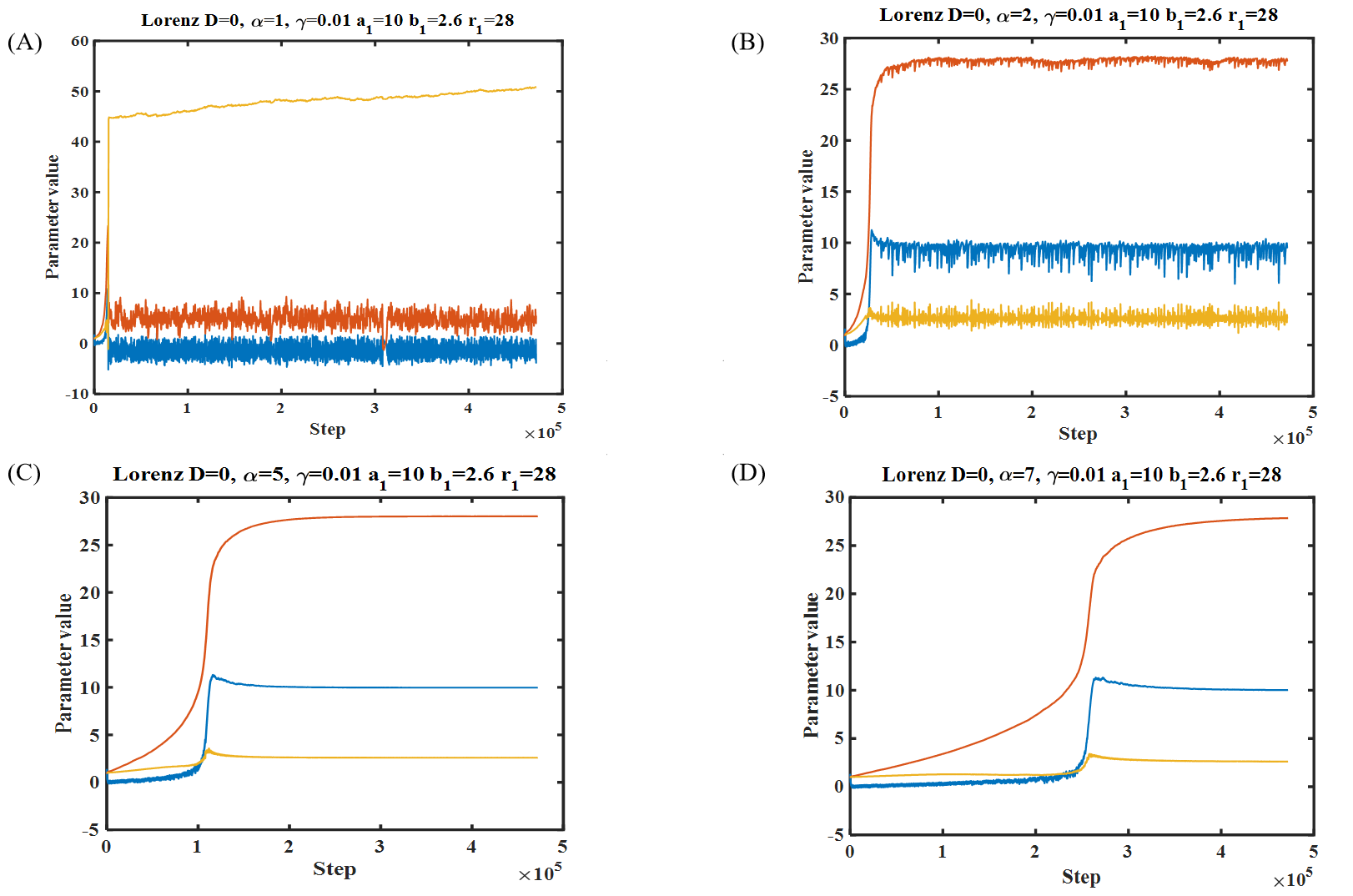}
	\caption{Reconstruction with different $\alpha$ at $t_ {unit} = 0.01,\gamma=0.01$, and $D = 0$. (A) $\alpha=1$. (B) $\alpha=2$. (C) $\alpha=5$. (D) $\alpha=7$.}
	\label{fig:f2}
\end{figure}

The Lorenz system is a famous deterministic dissipative system\cite{45lorenz1963deterministic}, which is written as
\begin{equation}
\label{eq:lorenz}
\begin{aligned}
&\dot{x}=a(y-x), \\
&\dot{y}=r x-y-x z, \\
&\dot{z}=x y-b z
\,.
\end{aligned}
\end{equation}
With the often-chosen parameters $a=10,b=2.6,r=28$, this system displays very typical chaotic behaviour with the maximum Lyapunov exponent close to $1$.
%\setlength{\abovecaptionskip}{-0.15cm}
%	\centering
%	\includegraphics[height=7cm,width=0.7\textwidth]{fig/fig0}
%	\caption{Time series of $x$ for the Lorenz system driven by white noise, where $t_ {unit} = 0.01,N_{step}=5000$, and the noise strength $D = 1$.}
%	\label{fig:f0}
%\end{figure}
If the time series of $x,y,z$ are all known, it is not hard to recover the coefficients of the polynomials on the right hand side of Eq.~(\ref{eq:lorenz})~\cite{232017Reconstruction,2016Discovering,2021Reconstruction,17Parlitz1996Estimating,34shandilya2011inferring}. Instead, if only one variable, say $x$, is observed, the recovery becomes much harder~\cite{41Wangsh}. Here, the new formulation is applied with two hidden variables $y,z$, three unknown parameters $a,r,b$ and two auxilliary parameters $e_2,e_3$. Thus, according to the above discussion, the total number of equations is $n_t=2\times(1+3+2)+(3+2)\times 2=22$. In the following simulations, $[a,r,b]=[1,1,1], [y,z]=[1,1]$ are conveniently used as the initial conditions for the corresponding variables. With the above algorithm, the three unknown parameters $a,r,b$ converge well as shown in Fig.~\ref{fig:f1}(A,B,C) and Fig. ~\ref{fig:f2}(C,D). For different hyper-parameters, the convergence seems to follow similar patterns. Initially, the unknown parameters change steadily but at some point start to rush to the true values and then remain there throughout the simulation. Nevertheless, the convergence is different in disparate conditions and could even go awry if improper hyper-parameters are chosen as displayed in Fig.~\ref{fig:f1}(D) and Fig. ~\ref{fig:f2}(A,B).  

In this example, the integration time step is $t_{\mbox{unit}}=0.01$. As can be seen from the figures, a total number of $N_{step}\sim 10^5$ steps are needed for the convergence, which is determined by the two hyper-parameters $\alpha\,,\gamma$. Currently, we take $\alpha=3$, indicating a memory of $1/\alpha=0.33$ which is about a third of the Lyapunov time of the system. The learning rate is chosen to be $\gamma=0.01$, about one hundredth of the Lyapunov exponent. $\gamma$ is used to control the update of the parameters in the evolution equation and should keep small compared with the time rate of the system state $(y,z)$ in order not to disrupt the generic dynamics of the original system.  As shown in Fig.~\ref{fig:f1}(A,B,C), a reduction of $\gamma$ extends the learning time and requires more data for convergence. Nevertheless, if the available time series has a limited length, a technique of data recycling may be employed for a successful reconstruction as shown in Appendix \ref{sec:app2}. On the other hand, if $\gamma$ is overly large, the parameters change rapidly and the original dynamics may be altered so that no convergence results, as displayed in Fig.~\ref{fig:f1}(D). So during the reconstruction, the learning rate should be chosen properly to ensure both the stability and efficiency.  

The decay rate $\alpha$ should not be too small in order to bound the integrals in Eq.~(\ref{eq:f4}). On the other hand, if $\alpha$ is too large, the memory becomes really short and these integrals highly depend on the current state which varies rapidly and easily overruns the slow parameter changes. In Fig.~\ref{fig:f2}, the influence of the decay rate $\alpha$ on the convergence is displayed. As $\alpha$ increases, the learning becomes slow and more data is needed, just like reducing $\gamma$. When $\alpha $ is relatively small(Fig.~\ref{fig:f2}(A)), the effective integration time becomes long and the integral will be very sensitive to parameter changes, causing instability in the system, unless the learning rate $\gamma$ is really small and averaging out the fast oscillations. Therefore, for a system with a large Lyapunov exponent, when the learning rate is fixed, it is necessary to apply a proper decay rate to reduce the sensitivity of the parameters (Supplementary Fig.~\ref{fig:alpha7}, Supplementary Fig.~\ref{fig:errorlya}). In general, it is important to adjust the learning and the decay rate to keep a balance between the two so as to achieve a stable and fast convergence.

\par
From the above discussions, it seems that the current reconstruction scheme is quite fragile. Actually, it is not and able to work quite well even in the presence of noise. Next, we investigate the impact of noise on the reconstruction since it inevitably exists in every experiment or observation. For this purpose, we add a noise term to each of the evolution equation~(\ref{eq:lorenz}), which describes white noise with the following characteristics.
\begin{subequations}
\label{eq:noise}
\begin{align}
  \label{eq:noise_m}<\eta_{i}(t)> &= 0 ,\\
  \label{eq:noise_d} <\eta_{i}(t)\eta_{j}(t')> &= 2D\delta_{ij}(t-t'),
\end{align}
\end{subequations}
where $\langle \cdot \rangle$ denotes an ensemble average and the delta function is a Kronecker one for the discrete index $i, j$, and a Dirac one for the continuous time variables $t-t^{\prime}$. $D$ is the noise intensity and when $D=0$ we recover the deterministic autonomous system. 

\begin{figure}[htbp]
	\centering
	\includegraphics[height=7cm,width=0.7\textwidth]{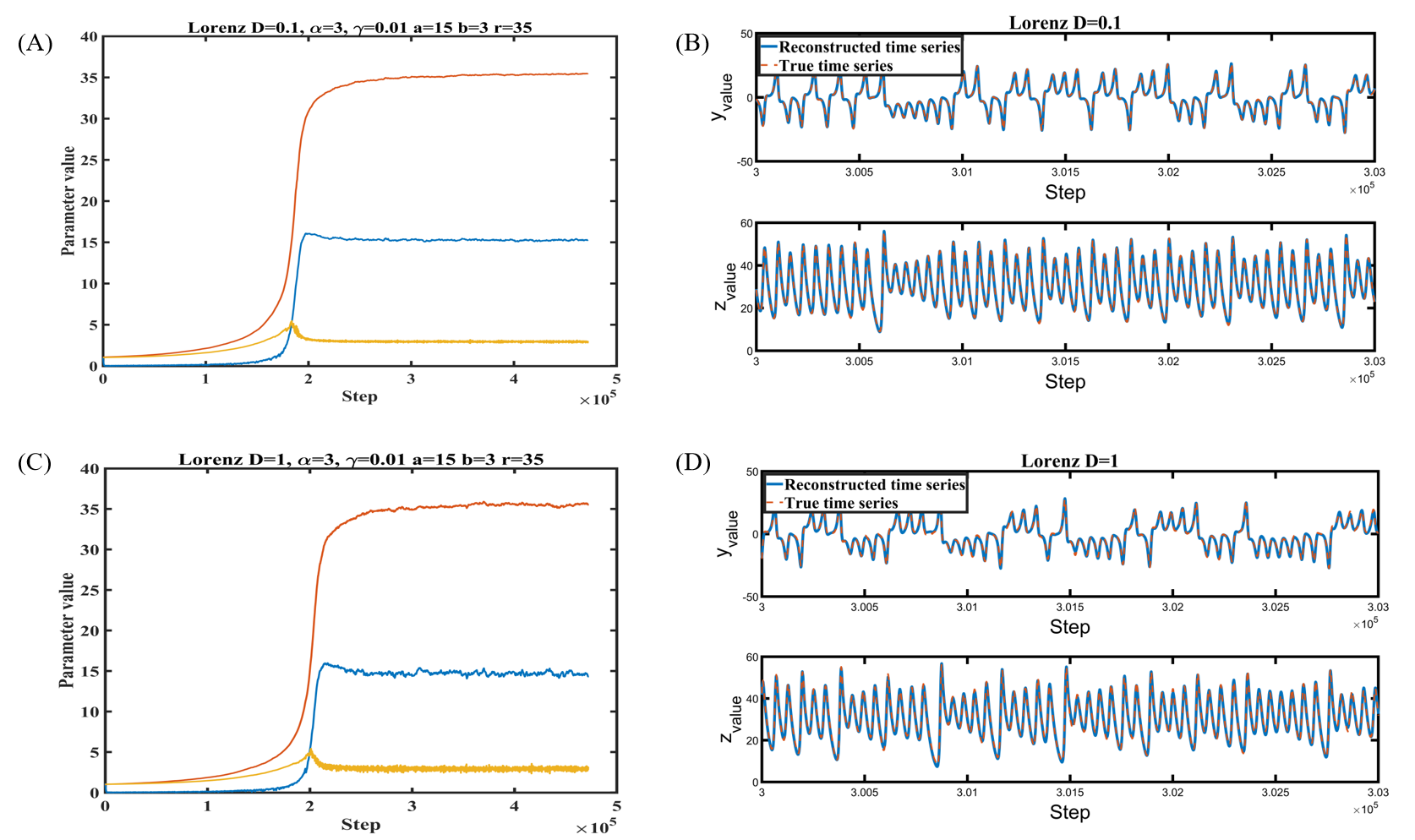}
	\caption{Parameter and trajectory construction at D=0.1 and D=1, with $t_ {unit} = 0.01,\gamma =0.01$, $\alpha =3 $. (A) The parameter evolution during the reconstruction. (B) The reconstructed time series of hidden variables $y$ and $z$ at D=0.1. (C) The parameter evolution during the reconstruction. (D) The reconstructed time series of hidden variables $y$ and $z$ at D=1.}
	\label{fig:f3}
\end{figure}

Even though a noisy orbit is generated with Eq.~(\ref{eq:lorenz}) with noise ($D\neq 0$), we still use the same deterministic formulation in the previous section. As an example, suppose that only the noisy $x-$orbit is given and we still aim to deduce the unknown parameters and the hidden variables $y(t),z(t)$. If the noise is not very strong, $x(t)$ may be directly substituted into the $22$ reconstruction evolution equations mentioned above, together with the $\dot{x}(t)$ computed with any fair smoothening technique~\cite{2021Reconstruction}. As a result, only the average orbit is reconstructed with the unknown parameters. Fig.~\ref{fig:f3} shows the reconstruction results at different noise intensities, {\em i.e.}, $D=0.1$ in Fig.~\ref{fig:f3}(A,B) and $D=1$ in (C,D). The convergence pattern of the parameters displayed in Fig.~\ref{fig:f3}(A,C) looks very similar to that in the deterministic case but small jittering exists due to noise perturbation, which is more clearly seen in (C) for $D=1$. After the small oscillations are averaged out, we get a fairly good estimate of the parameters, which is listed in Table ~\ref{tab:table1}. In fact, the results of extra tries are also displayed in the Table up to $D=5$. In general, the estimation gets worse for greater noise intensity as expected but the errors do not appear monotone. The reconstructed orbits $y(t),z(t)$ are displayed in Fig.~\ref{fig:f3}(B,D), which seem almost overlapping with the true orbits, just as in the deterministic case. 

From the above analysis, it can be seen that the new construction scheme is quite robust against noise as long as it is not too strong. Next, the scheme will be applied to several typical examples in nonlinear dynamics with high dimensions to show its validity.

\begin{table}[htb!]
	\centering
	\caption{\label{tab:table1}The dependence of parameter reconstruction on noise in the Lorenz system.}
	\begin{tabular}{c|cccc}
		\hline
		       & $D=0$ &$D=0.1$ &$D=1$ &$D=5$\\
		\hline
		$r=35$ & $r=35.19$ &$r=35.35$ &$r=35.52$ & $r=34.56$\\
		$b=3$  & $b=2.95$  &$b=2.94$  &$b=2.93$  & $b=3.07$\\
		
		$a=15$ & $a=15.26$ &$a=15.26$ &$a=14.68$ & $a=13.55$\\
		\hline
	\end{tabular}
\end{table}
\subsubsection*{The $N$ coupled Lorenz oscillators}
First, we consider the inference problem in a network with $N$ nodes each of which is occupied by a Lorenz system. The connection between nodes marks their interaction. More specifically, the $i-$th node is governed by the equation below 
\begin{align}
  \dot x_{i} &=a_{i}(y_{i}-x_{i})+\eta_{i}(t),\notag \\
  \dot y_{i} &=r_{i}x_{i}-y_{i}-x_{i}z_{i}+\sum_{j=1}^N A_{ij}(y_{j}-y_{i})+\eta_{i}(t), \notag  \\
  \dot z_{i} &=x_{i}y_{i}-b_{i}z_{i}+\eta_{i}(t)\,
  \label{eq:f19}
\end{align}
where $\eta_i(t)$ is the above-mentioned Gaussian white noise and $A_{ij}$ is the interaction matrix. Here, without less of generality, the interaction is assumed to take place between $y-$components. The local parameters are randomly selected from  intervals on the positive real axis: $a_{i}\in [10,12], r_{i}\in [28,30], b_{i}\in [2.6,2.8]$. The noise intensity $D=0.01$ and the sampling interval $t_ {unit} = 0.01$.  Suppose that only the $x-$components are measured.  Our goal is to reconstruct the coupling matrix $A_{ij}$ and all the local parameters $\{a_i,b_i,r_i\}_{i=1,2,\cdots,N}$ as well as the time-dependent hidden variables. 

\begin{figure}[htbp]
	\centering
	\includegraphics[height=7.85cm,width=0.75\textwidth]{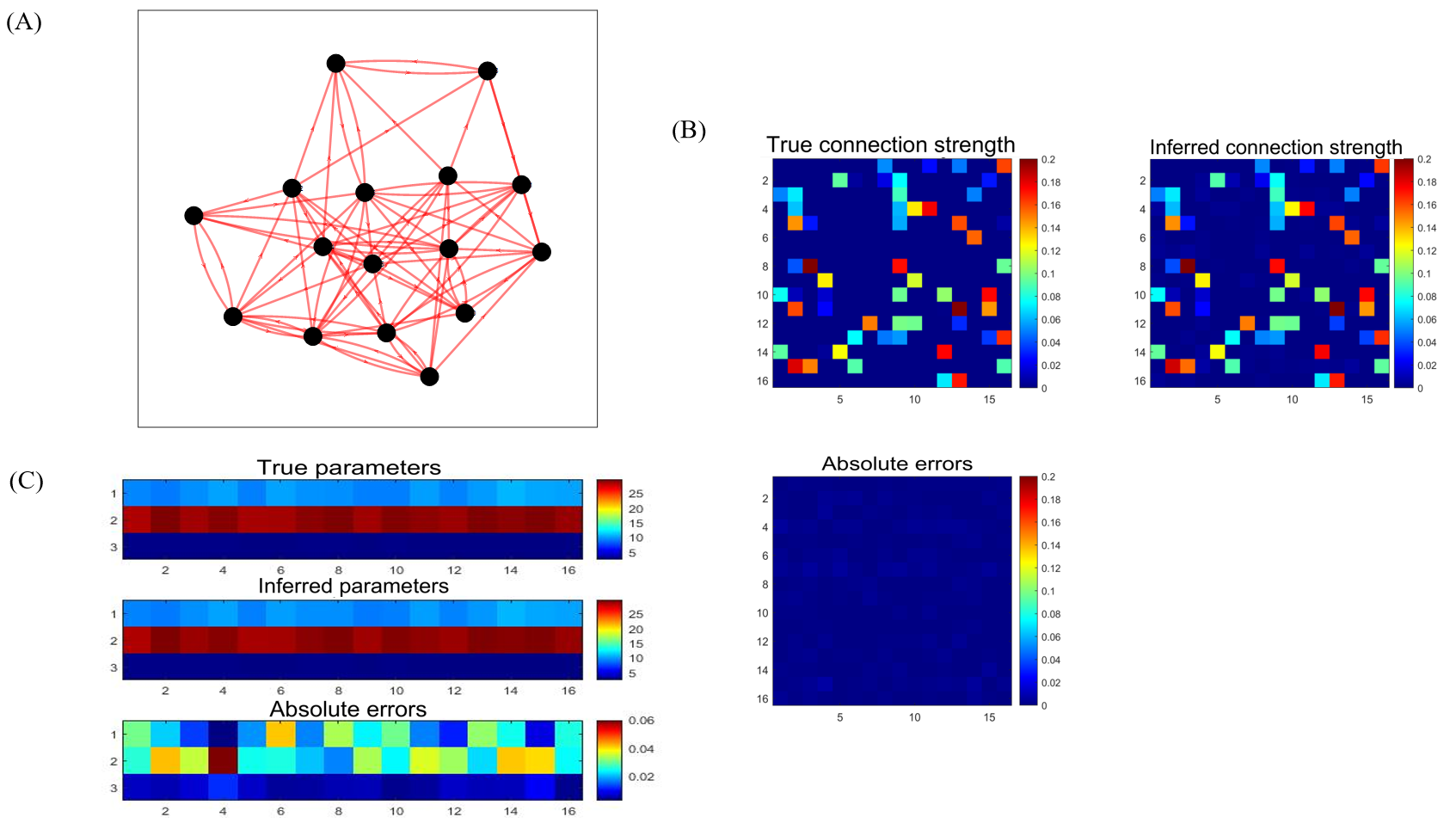}
	\caption{Reconstruction in a network of 16 coupled Lorenz systems with different set of parameters, where $A_{i j}\in[0,0.2]$,$\alpha=3$,$\gamma=0.01$.  (A) The coupling topology between the 16 Lorenz systems. (B) The actual and inferred network coupling strengths together with the absolute inference errors.  (C) The nod actual and inferred local node parameters and the absolute errors.}
	\label{fig:f5}
\end{figure}

With the above scheme, for an \text { Erdős-Rényi (ER)}  network with $N=16$, the coupling matrix and the node parameters are correctly deduced as displayed in  Fig.~\ref{fig:f5}. For simplicity, we take the initial conditions of interaction matrix $A_{ij}$ to be zero unless otherwise stated,  and the initial values of local parameters are $[a_{i},r_{i},b_{i}]=[1,1,1], [y_{i},z_{i}]=[1,1]$. After the reconstruction, the absolute errors of the computation in the coupling matrix are of the order of $10^{-3}$, and around $10^{-2}$ in the node parameters. The time series of all hidden variables are also reconstructed at the same time, which is not shown here for brevity. Please check Fig.~\ref{fig:lorenn=8} in the Supplementary for one example.

When $N$ is small, we may start with a full matrix $[A_{i,j}]$ containing $N^2$ elements to be determined. The network structure can be drawn from the non-zero elements of $[A_{i,j}]$ after the reconstruction and we do not need any prior information about the network topology. Nevertheless, this type of analysis is not scalable since the quadratic increase of the number of the unknown matrix elements with the network size prevents their efficient estimation based only on the linearly increasing information. In fact, the current brute-force method works up to $N \sim 20$. For larger networks, extra information about the network topology helps a lot, especially those that cut down the number of unknown matrix elements. For example, if the structure of the network is approximately known (which may still contain some spurious links), our computation remains valid as long as the number of links increases not that fast. 

\begin{figure}[htbp]
	\centering
	\includegraphics[height=8cm,width=0.75\textwidth]{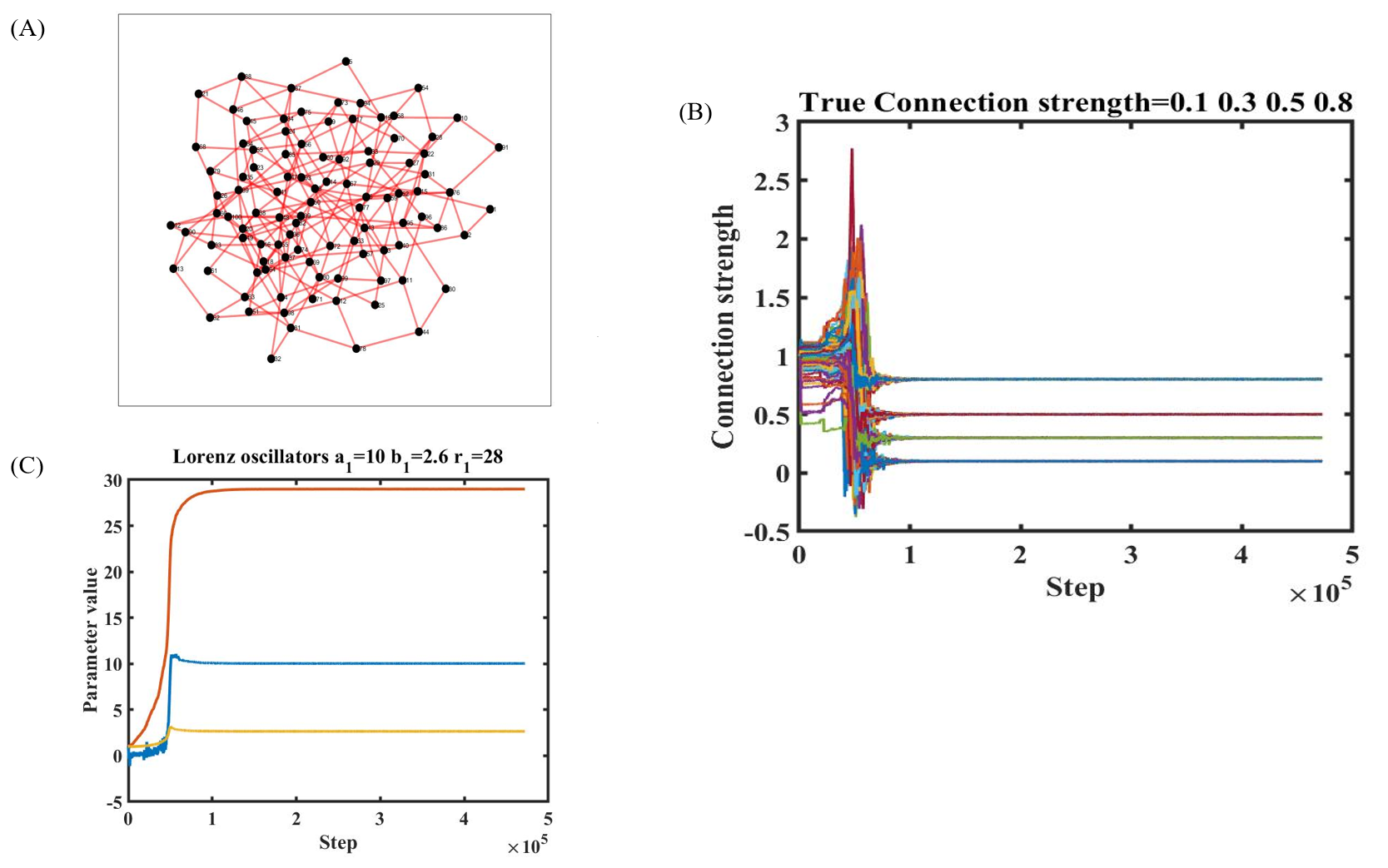}
	\caption{A small-world network (N=100) with an average degree of 4, with $[a_{i},r_{i},b_{i}]=[1,1,1]$, $[y_{i},z_{i}]=[1,1]$,$\alpha=3$,$\gamma=0.01$. (A) The coupling network of 100 Lorenz systems. Inference of the network coupling strength (B) and the parameters of a node (C).}
	\label{fig:f6}
\end{figure}

Here, we tried the reconstruction on a small-world network with $N=100$ and an average degree $K=4$ as portrayed in Fig.~\ref{fig:f6}(A). For an easy comparison, the non-zero coupling is allowed taking only four values $0.1,0.3,0.5,0.8$.  The results are displayed in Fig.~\ref{fig:f6}(B,C) which agree well with the original values. The convergence of the local node parameters looks quite similar as in precious cases (see Fig.~\ref{fig:f6}(B)). However, the evolution of the couplings starts more erratically. The fluctuations grow larger and larger and culminate just before a sudden convergence to the true values (see Fig.~\ref{fig:f6}(C)). The whole process closely resembles what happens in chaos control: only when the system state gets close to the stable manifold of the true dynamics and the true parameters does the gradient descent scheme in Eq.~(\ref{eq:f8}) start to work and drive the wandering point to the right solution\cite{ottChaos}. 
%%%%%%%%%%%%%%%%%%%%%%%%%%%%%%%%%%%%%%%%%
%%%%%%%%%%%%%%%%%%%%%%%%%%%%%%%%%%%%%%%%

%%%%%%%%%%%%%%%%%%%%%%%%%%%%%%%%%%%%%%%
%%%%%%%%%%%%%%%%%%%%%%%%%%%%%%%%%%%%%%%%%%%%%%%%%%%%%%%%%

In the current example, we have $700$ unknown system parameters, including $300$ node parameters and $400$ coupling strengths. In addition, $200$ auxiliary parameters $\mathbf{e}$ are needed to drive $y_i(t)\,,z_i(t)$ to the right trajectories. Altogether, we have $900$ hundred parameters, which will generate over $180,000$ equations in the new formulation, if we exactly follow the procedures in the theory part. However, a convenient approximation could be adopted based on the given network structure and the observation that the influence of a node on other nodes decays quickly in an asynchronous state. According to the equation of motion, each of the above-mentioned parameters could be assigned to a node or a connection between nodes. To the lowest order, it is reasonable to assume that the dynamics at a node mainly depends on the parameters of neighbouring nodes or edges. With this assumption, the partial derivatives with parameters are taken to be zero if they locate outside the immediate neighbourhood. As a result, the number of equations is greatly cut down, which in fact grows only linearly with the size of the network with a fixed average degree. 

\par
\begin{figure}[!htbp]
	\centering
	\includegraphics[height=5.5cm,width=0.7\textwidth]{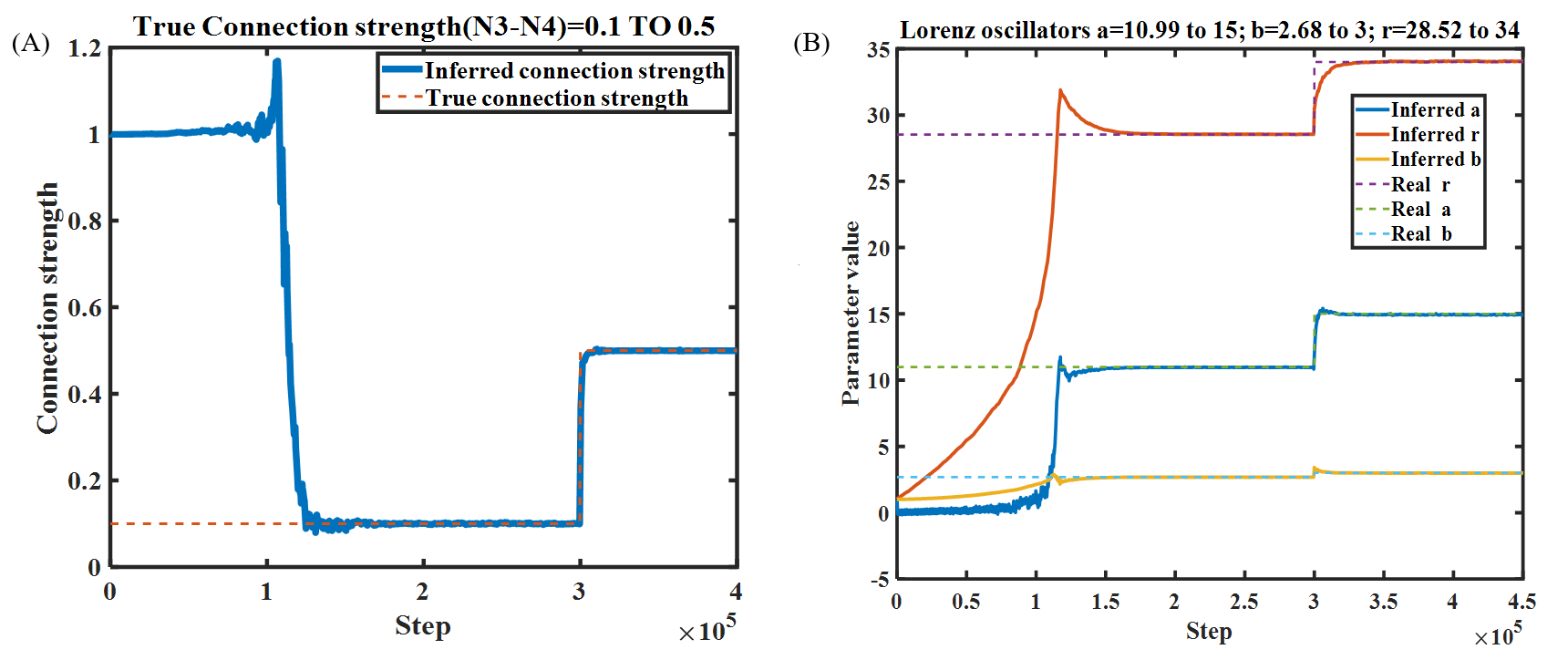}
	\caption{Monitoring changes in network topology and node dynamics. (A) The change in coupling strength between nodes over time. (B) The change of parameters in node dynamics over time.}
	\label{fig:f4_4_4}
\end{figure}

As the true values of the parameters or state variables are obtained through a co-evolution with the observed time series, the proposed method can be applied online to monitor changes in network topology or node dynamics if the change does not happen so frequently. As an example, the above formulation is directly applied to the network with $8$ nodes displayed in Fig.~\ref{fig:lorenn=8}(A) in the Supplementary. Starting from a somewhat arbitrary initial guess value $1$, the coupling converges to the true value $0.1$ after a quite long transient. Later on, as the coupling is switched to $0.5$ in the network, the reconstruction algorithm detects and quickly follows this change. It is interesting to see that the reaction is really fast this time and the transient is almost unnoticeable as shown in Fig.~\ref{fig:f4_4_4}(A). The node parameters that change at the same time can be similarly observed, as shown in Fig.~\ref{fig:f4_4_4}(B). One reason for the quick response for the parameter change is that the state variables $\{y_i(t),z_i(t)\}_{i=1,2,\cdots,N}$ remain close to the true values in this process so that the convergence seems monotone and exponential!  
\par

\subsubsection*{The FitzHugh-Nagumo networks}
A salient feature of the FitzHugh-Nagumo (FHN) system is the excitability - once the potential exceeds a threshold, the system state immediately shoots to a very high value and then relaxes back to the base. Therefore, the temporal pattern is quite different from that of a Lorentz system. Is it still possible to apply our method to this new situation? Below, we check the continued validity of the current scheme.

The FHN equation is a second-order nonlinear system used to simulate the dynamics of a neuron. To model neural networks, these neurons connect to each other and interact with their neighbours. The equations of motion can be written as
\begin{equation}
\label{FHNEQ}
\begin{aligned}
&\dot{v}_{i}=v_{i}-v_{i}^{3}-w_{i}-\sum_{j=1}^{N} A_{i j}\left(v_{j}-v_{i}\right)+\Gamma_{1,i} \\
&\dot{w}_{i}=a_{i}+b_{i}+c_{i} w_{i}+\Gamma_{2,i}
\end{aligned}
\end{equation}
where $i=1,2,\cdots,N$ denotes different neurons, $v_{i}$ is the rate of change of the membrane potential, $w_{i}$ is the slowly changing recovery variable of the $i$-th neuron and $N$ is the number of neurons. In this model, the dynamical behaviour of the system largely hinges on the coefficients $a_{i}\in[0.3,0.5],b_{i}\in[0.5,0.7],c_{i}\in[-0.04,-0.03]$ and the interaction weight $A_{i j}\in[0,1]$. $\Gamma_{1,i}, \Gamma_{2,i}$ are noise terms of the two components, which are usually taken to be white satisfying Eq.~(\ref{eq:noise}), with $D=0.01$.

\begin{figure}[htbp]
	\centering
	\includegraphics[height=7.5cm,width=0.75\textwidth]{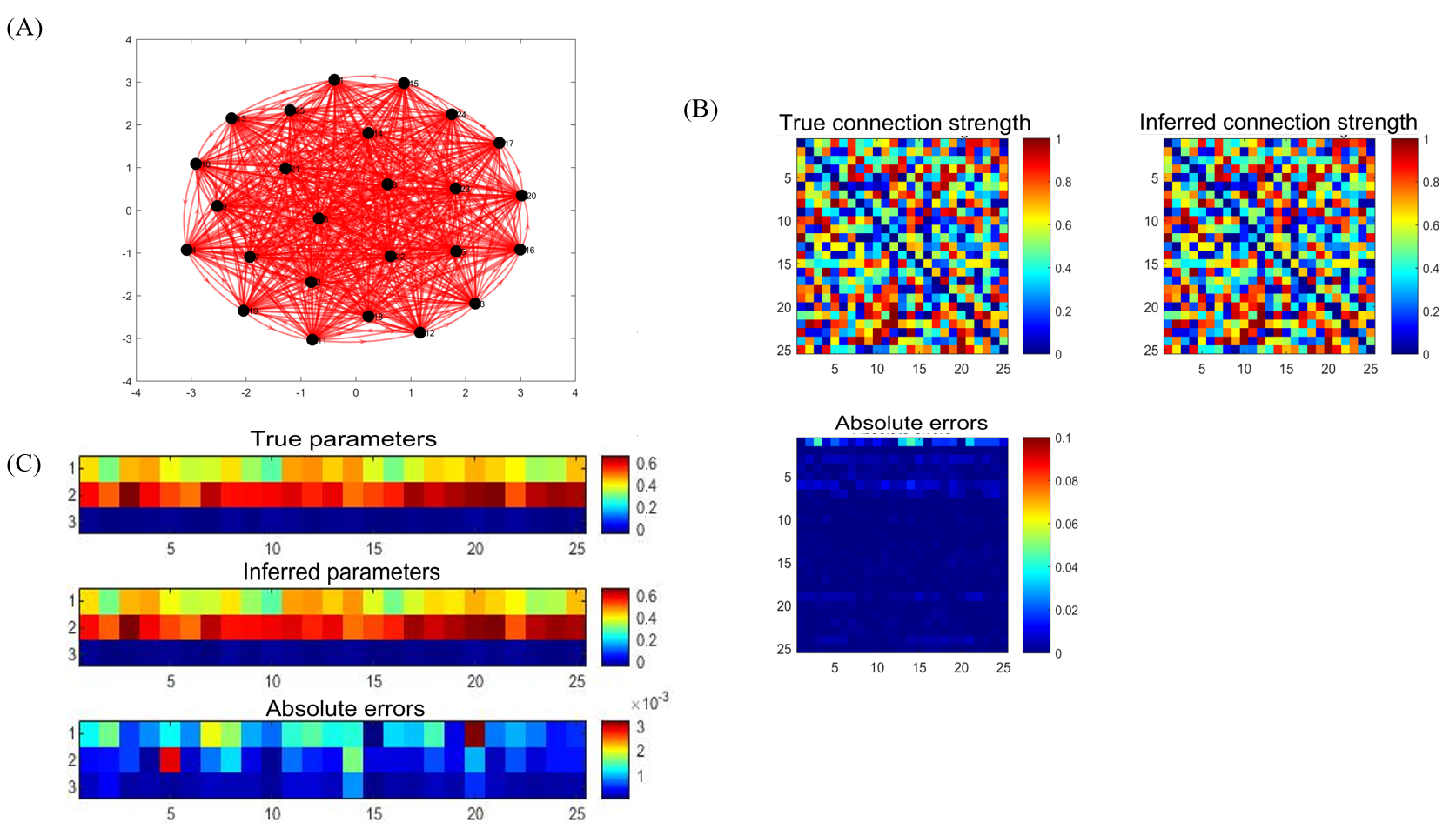}
	\caption{Reconstruction in a network of $25$ FHN nodes with $\gamma =0.5$, $\alpha =5 $. (A) The network topology. (B) The actual and inferred network coupling strengths and the absolute errors.  (C) The actual and inferred local parameters and the absolute errors.}
	\label{fig:f8}
\end{figure}

\begin{figure}[htbp]
	\centering
	\includegraphics[height=8cm,width=0.7\textwidth]{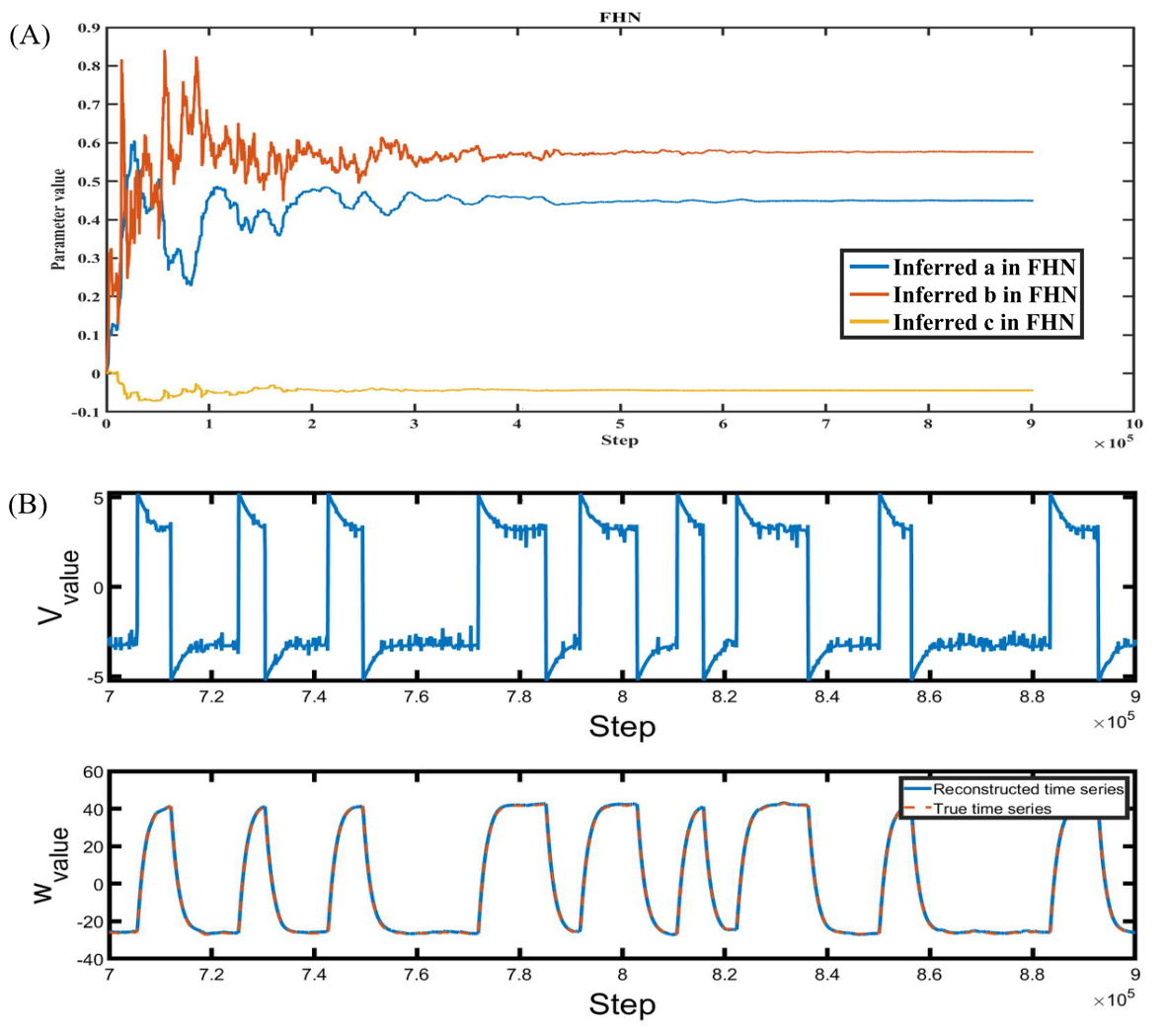}
	\caption{The parameter reconstruction and the evolution of state variables at one node of the FHN network. (A) The parameter reconstruction. (B) The measured time series of $v$ and $w$ together with the reconstructed $w$.}
	\label{fig:f9}
\end{figure}

In practice, only the potential $v_{i}(t)$ of each neuron is subject to a direct measurement. Is it possible to recover $w_{i}(t)$ and all the parameters in the equation with this information?  Fig.~\ref{fig:f8}(A) displays a \text { Erdős-Rényi (ER)} network with $N=25$ from which the measured data was obtained. As usual, we apply the above scheme by initially setting all unknown coupling strength $A_{ij}$ to $1$ and  the local system parameters to $0$. With this network size, all the parameters could be obtained directly without other assumptions. The obtained coupling strengths are depicted in  Fig.~\ref{fig:f8}(B) with the absolute error of the order of $10^{-2}$, while the local parameters are portrayed in Fig.~\ref{fig:f8}(C) with the absolute error of the order of $10^{-3}$. Fig.~\ref{fig:f9}(A) shows the convergence of local parameters for a node from the initial values $0$ to the correct values, where the oscillatory character at the beginning is still clearly seen. Fig.~\ref{fig:f9}(B) displays the time series of $v_{i}(t)$ where the profile is contaminated partly by visible noise. However, the reconstructed time series of $w_{i}$ seems smooth and overlaps with the true one almost perfectly.
%%%%%%%%%%%%%%%%%%%%%%%%%%%%%%%%%%%%%%%%%%%%%%%%%%%
%%%%%%%%%%%%%%%%%%%%%%%%%%%%%%%%%%%%%%%%%%%%
%%%%%%%%%%%%%%%%%%%%%%%%%%%%%%%%%%%%%%%%%%%%

\begin{figure}[htbp]
	\centering
	\includegraphics[height=9cm,width=0.7\textwidth]{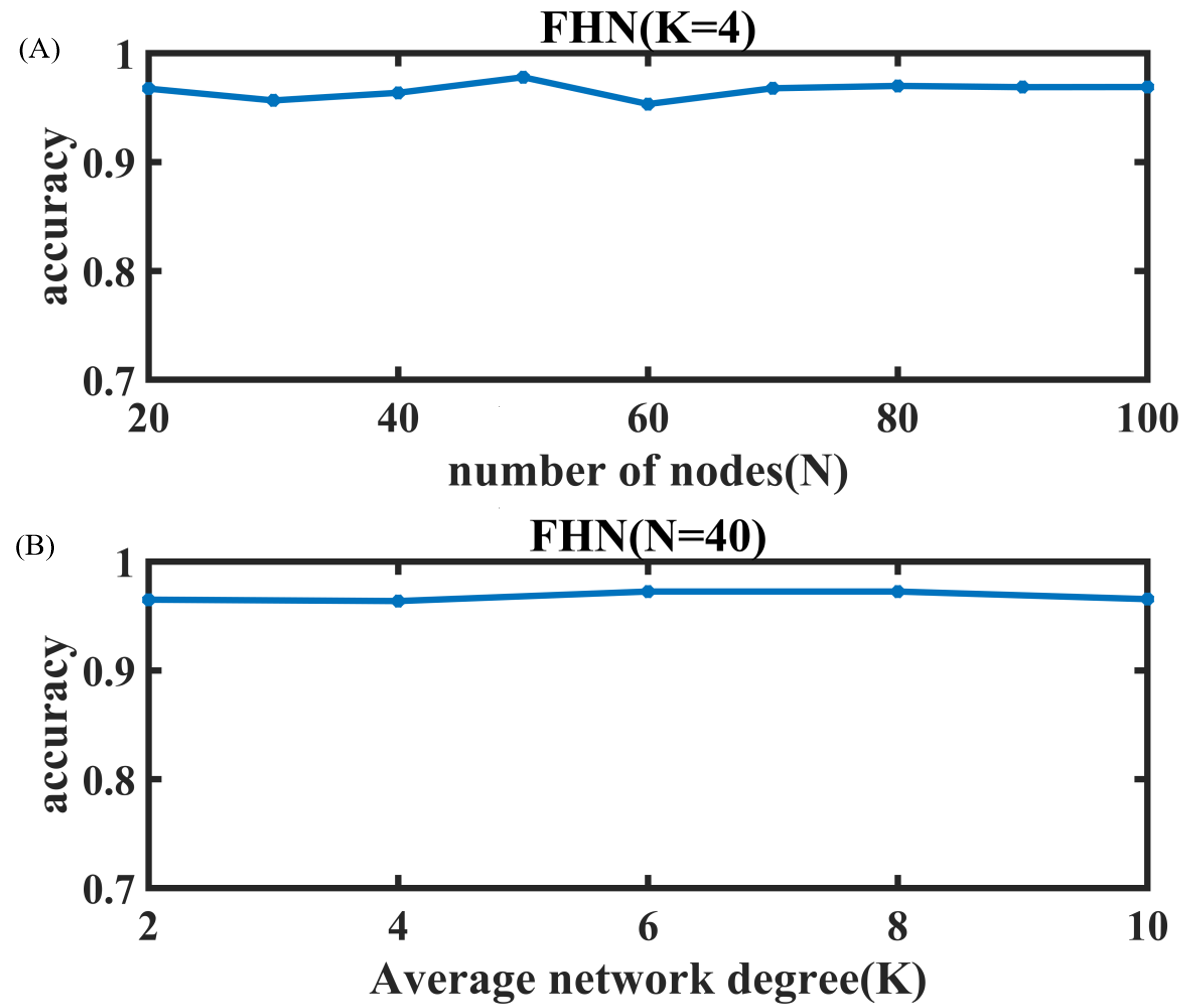}
	\caption{How the accuracy of network reconstruction depends on (A) the network size and (B) the average degree.}
	\label{fig:f10}
\end{figure}

For small-world FHN networks with known structures, for example, the one in Fig.~\ref{fig:f6}(A), how does the reconstruction accuracy depend on the network size? If the average degree is fixed to $4$, this dependence is shown in Fig.~\ref{fig:f10}(A). It is easy to see that the inference accuracy can be maintained above $95\%$ after averaging over 20 realizations. The accuracy is defined as\cite{34shandilya2011inferring}
\begin{equation}
Ac_{\rho}=\frac{1}{K \times N} \sum_{i, j} H\left((1-\rho)-\Delta A_{i j}\right)
\,,
\label{eq:fac}
\end{equation}
where $\rho$ is the required accuracy and we take $\rho=0.1$ here. $H$ is the Heaviside function and the relative error between the reconstructed ($\hat{A}_{i j}$) and the real coupling ($A_{i j}$) is:
\begin{equation}
\Delta A_{i j}=\left|\hat{A}_{i j}-A_{i j}\right| /\left(A_{i j}\right)
\,.
\end{equation}

The main source of error is the synchronization among part of the nodes, which makes it hard to distinguish them and thus leads to incorrect inferences. More discussion will be made below. On the other hand, if we keep the network size and increase the average degree, the inference will become harder since more couplings need deducing. Nevertheless, according to Fig.~\ref{fig:f10}(B), the accuracy still maintains near the value $95\%$ throughout. 
%%%%%%%%%%%%%%%%%%%%%%%%%%%%%%%%%%%%%
%%%%%%%%%%%%%%%%%%%%%%%%%%%%%%%%%%%%%%%%%
\subsubsection*{The Kuramoto oscillators}
The ability to deduce a large number of parameters from the time series may be applied to the reconstruction of networks with hidden nodes, about which little information is known. Even their existence has to be deduced from the measurement on neighbouring nodes as done in previous works to determine the possible connections of hidden nodes to the known ones in the network~\cite{292012Detecting,302014Uncovering,31Ri2016Data,322014Reconstructing}. Nevertheless, this type of deduction is able to confirm the existence of hidden nodes but the precise connections may not be inferred very accurately. In our scheme, all possible links may be included in the deduction as long as the node's existence is expected. The spurious ones will give a weight near zero and the weights will be recovered for the true ones. Also, the dynamics and internal parameters of the hidden nodes could be deduced as well. 

Here, Kuramoto oscillator model\cite{Kuramoto} defined on networks will be used as an example for this computation.  Each oscillator in the model is described with a phase, so $D=1$. The coupling between them induces synchronization at various levels, which has been studied comprehensively\cite{PIETRAS20191,Acebrn2005TheKM}. Consider a system with $N$ oscillators
\begin{equation}
\dot{\theta}_{i}=\omega_{i}+\frac{K}{N} \sum_{j=1}^{N} a_{i j} \sin \left(\theta_{j}-\theta_{i}\right),(i=1, \cdots, N)
\,,
\end{equation}
where $\theta_{i}$ and $\omega_{i}$  are the phase and the natural frequency of the $i$th oscillator, $K>0$ is the coupling strength and $a_{ij}$ is the adjacency matrix of the network. When node $i$ and node $j$ are connected, $a_{i j}=1$; otherwise $a_{i j}=0$. 

To characterize synchronization, in the literature~\cite{Acebrn2005TheKM}, an order parameter $R$ is defined as follows
\begin{equation}
R e^{i\psi}=\frac{1}{N}\sum_j e^{i \theta_j}
\,, \label{eq:order}
\end{equation}
where $R,\psi \in \mathbb{R}$. If $K\ll 1$, each oscillator rotates near its natural frequency and $R\sim 0$. With increasing $K$, the average frequencies of connected oscillators approach the average natural frequency in general and some of them may start to rotate together, reaching a partial synchronization such that $R>0$. For sufficiently large $K$, all the oscillators become synchronous and rotate with a common frequency $\sum_i \omega_i/N$, so that $R \sim 1$. During the course of dynamics or parameter deduction, synchronization is harmful as seen previously and will be explained below.   
\par
\begin{figure}[htbp]
	\centering
	\includegraphics[height=8.8cm,width=0.8\textwidth]{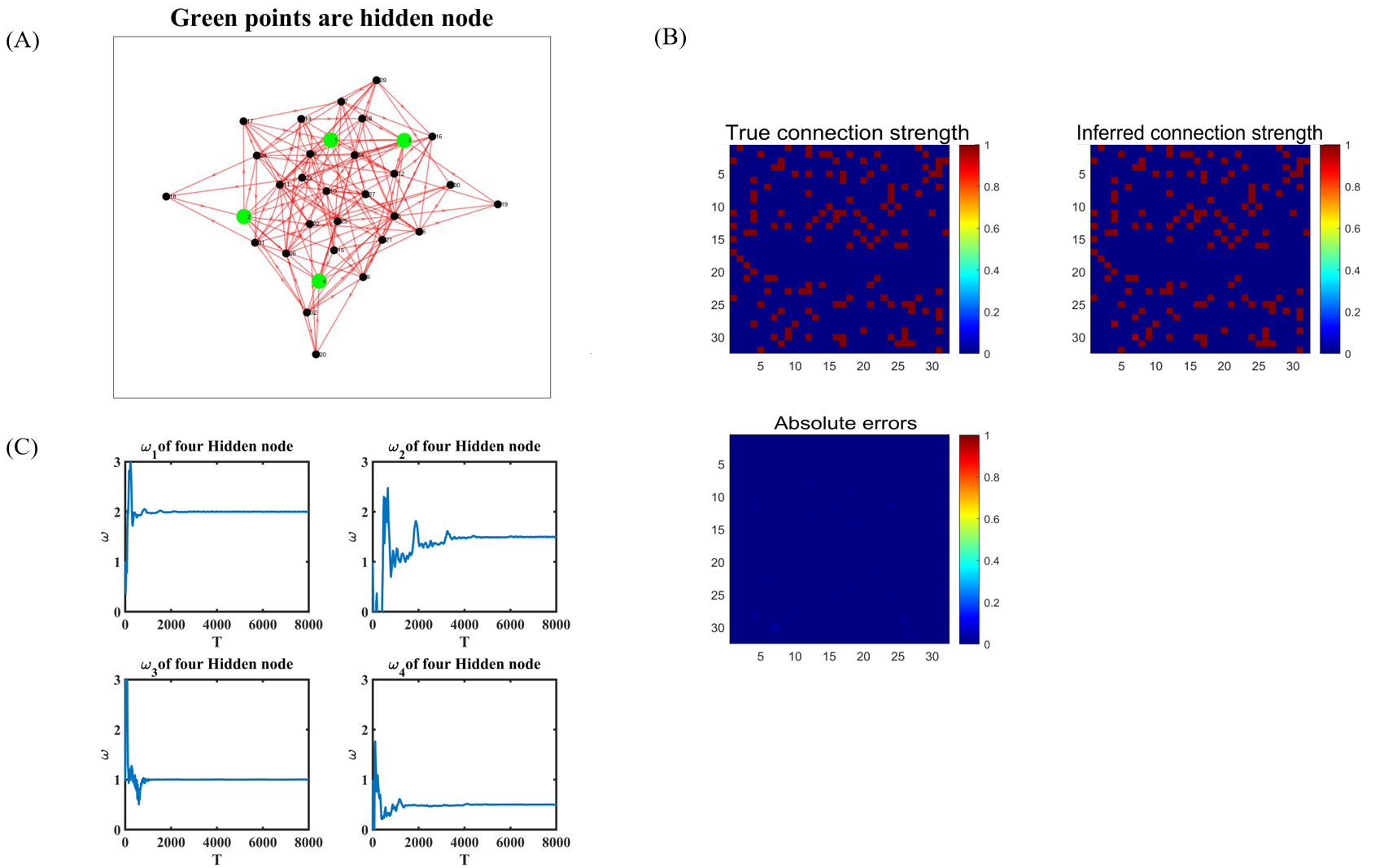}
	\caption{The parameter estimation of a network of Kuramoto oscillators, with $\gamma =0.5$, $\alpha =5 $.  (A) The Kuramoto oscillator network, where the green dots are hidden nodes. (B) The actual and inferred network coupling strengths, and the corresponding absolute errors.  (C) Inference of natural frequencies of hidden nodes.}
	\label{fig:fig11}
\end{figure}

In the current example, $32$ oscillators are connected to form an \text { Erdős-Rényi (ER)} network as shown in Fig.~\ref{fig:fig11}(A) and $\omega_{i}$'s are picked up from a uniform distribution in the interval $[0,20]$. The coupling strength $K=1$ is not very large to prevent synchronization. We assume that four hidden nodes exist in the network. Their dynamics and natural frequencies are not revealed to us, while the time series of the remaining nodes are known. First, the very existence could be deduced together with their rough connections with existing methods in the literature~\cite{292012Detecting}. Next, the whole network structure and the natural frequencies of these hidden nodes are reconstructed with the formalism introduced in this manuscript. In the following, the hidden nodes are randomly selected from the network with no neighbouring pair. 

\begin{figure}[htbp]
	\centering
	\includegraphics[height=8cm,width=0.7\textwidth]{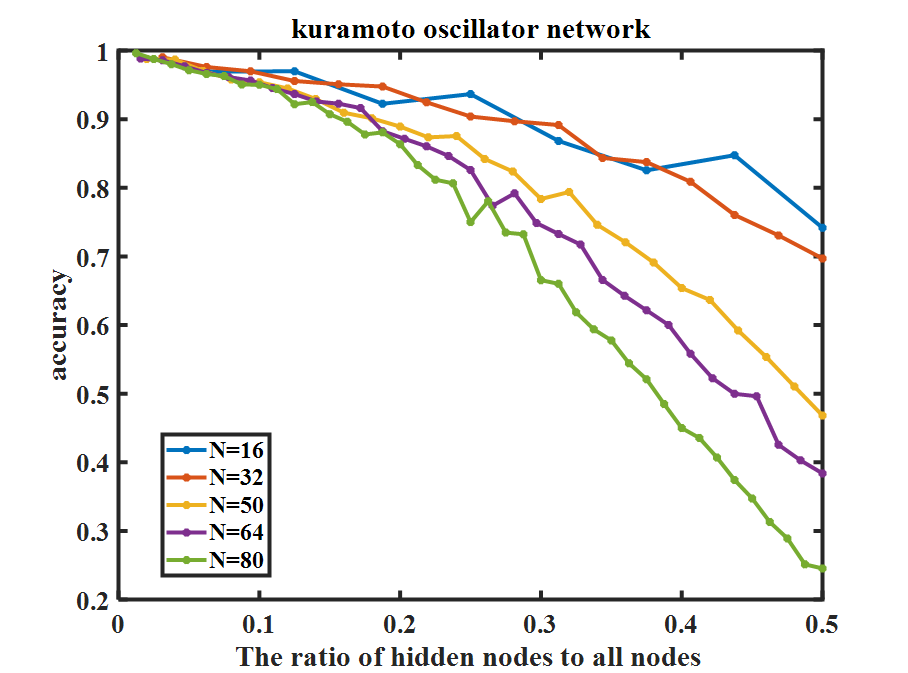}
	\caption{The inference accuracy vs the proportion of hidden nodes for different network sizes.}
	\label{fig:fig12}
\end{figure}

In this example, as the number of nodes is not large, as long as the very existence of hidden four nodes are known, we may start with a full connection matrix in the reconstruction. That is, the initial values of all entries of $[a_{ij}]$ are assumed to be $0.5$,and the natural frequency $\omega_{i}$'s start with the value $1$. In Fig.~\ref{fig:fig11}(B), a comparison is made between the simulation results and true values of the natural frequencies. The error seems small, not exceeding a few percent. The convergence of the natural frequencies of the hidden nodes is displayed in Fig.~\ref{fig:fig11}(C). The familiar oscillatory profiles appear again but they decay much faster than in previous examples. In Fig.~\ref{fig:fig12}, it is easy to see that the accuracy of the reconstruction (see Eq.~(\ref{eq:fac})) decreases with the increase of hidden nodes. The larger the network, the faster the decay. Nevertheless, from the figure we see that when the hidden nodes are less than $15\%$, no matter how large the network size is, the accuracy is higher than $90\%$. 

\par
\begin{figure}[htbp]
	\centering
	\includegraphics[height=8.5cm,width=0.7\textwidth]{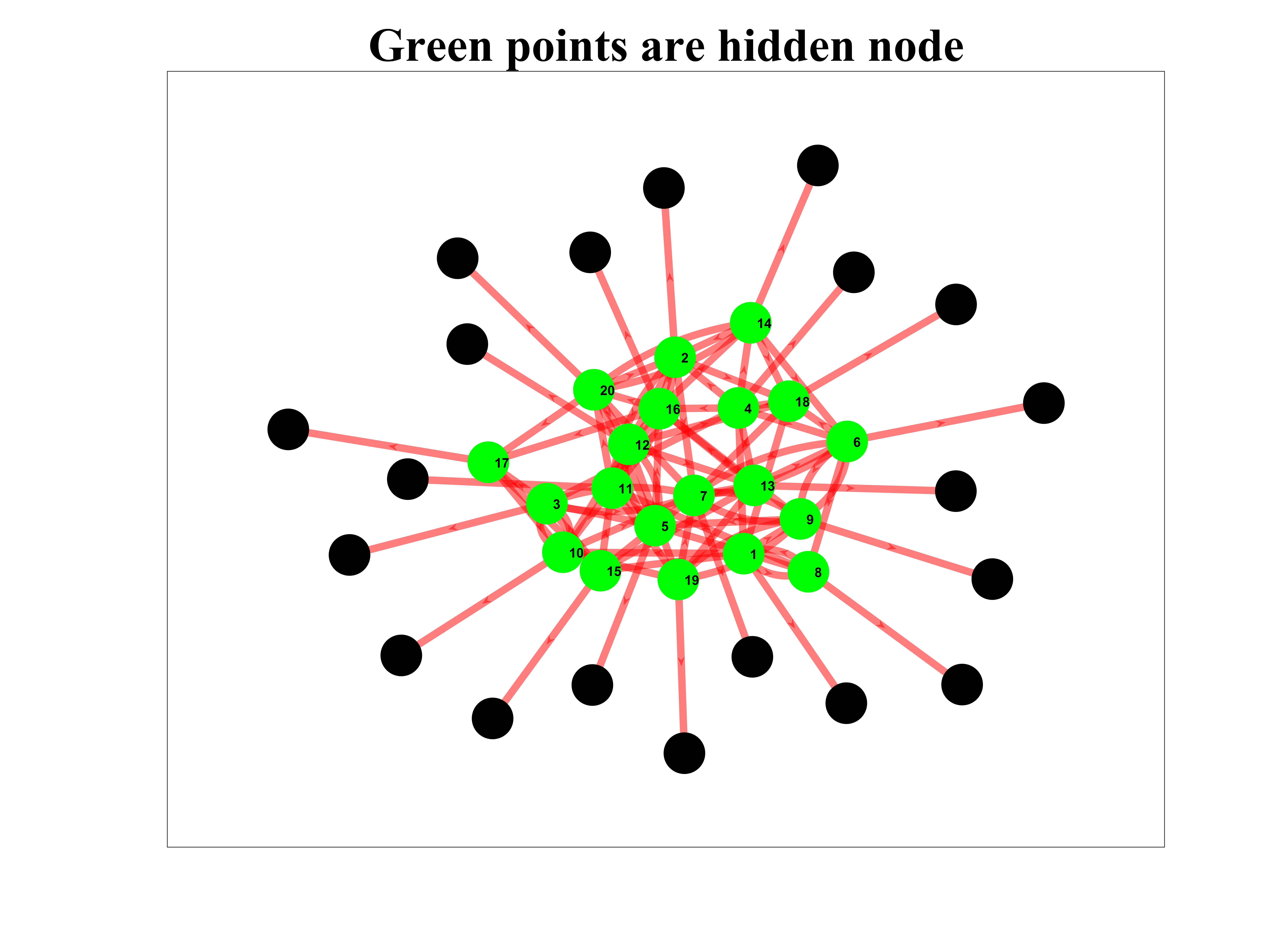}
	\caption{Inference of the structure of a cluster. The green points in the figure are hidden nodes, each of which is connected to a measurable node (black).}
	\label{fig:yintuan_shiytu}
\end{figure}
\begin{figure}[htbp]
	\centering
	\includegraphics[height=6.5cm,width=0.9\textwidth]{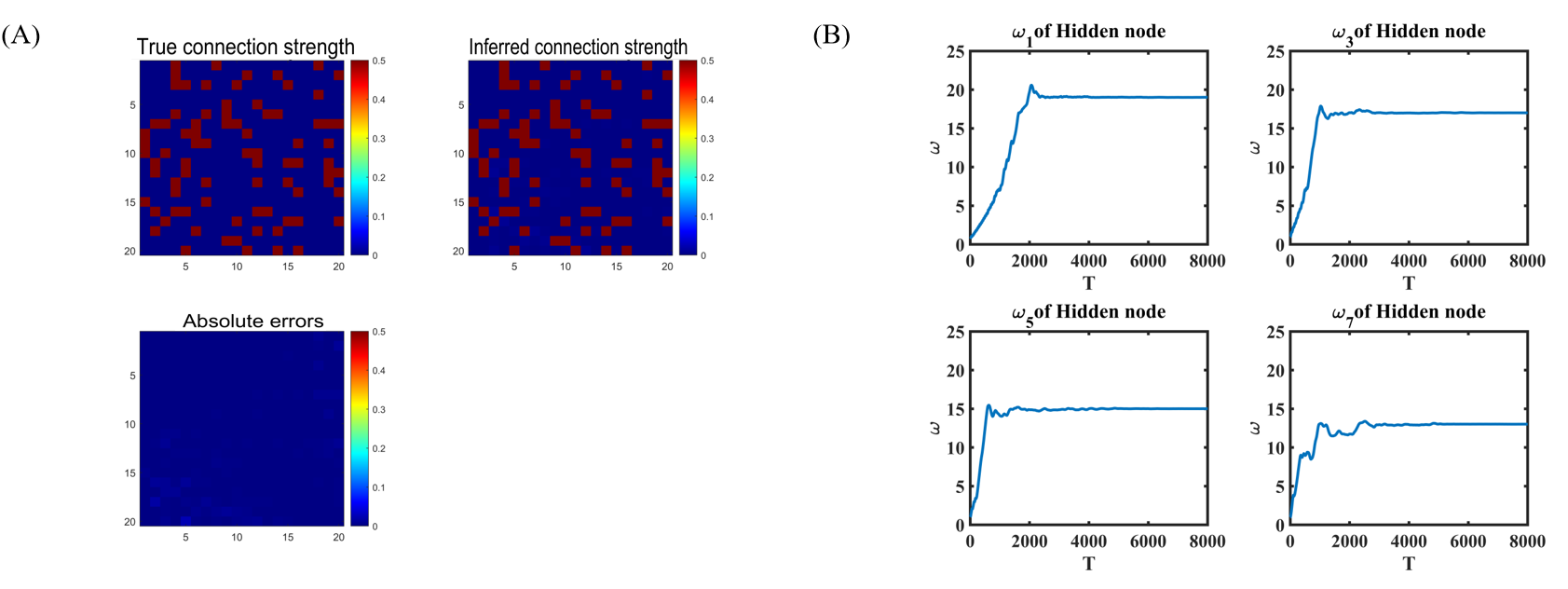}
	\caption{The cluster inference results. (A) The actual and inferred network coupling strength, and corresponding absolute errors.(B) Convergence of the natural frequencies of four hidden nodes in the hidden cluster.}
	\label{fig:yin_k}
\end{figure}

In fact, our method is even able to infer the structure of a cluster of hidden nodes. In Fig.~(\ref{fig:yintuan_shiytu}), there are 20 hidden nodes (green dots) forming a small cluster. The structure as well as all the parameters is unknown and none of them is observed directly.We can only measure their neighbouring nodes (black ones). From the measured time series, the natural frequencies of nodes in the hidden cluster and the coupling strengths among nodes can be inferred quite accurately, which is clearly displayed in Fig.~\ref{fig:yin_k}(A). A full connection between hidden nodes is assumed at the start and the algorithm will computed the actual strengths of the links: about $0$ for the non-existing ones and $0.5$ for the true connections. The convergence seems fast (see Fig.~\ref{fig:yin_k}(B)).
%%%%%%%%%%%%%%%%%%%%%%%%%%%%%%%%%%%%%%%%%%%%%%%%%%%%%
%%%%%%%%%%%%%%%%%%%%%%%%%%%%%%%%%%%%%%%%%%%%%%%%%%%%%%%%%%%%
%%%%%%%%%%%%%%%%%%%%%%%%%%%%%%%%%%%%%%%%%%%%%%%%%%%%%%%%%%%
\subsubsection*{Synchronization undermines deduction}
\begin{figure}[htbp]
	\centering
	\includegraphics[height=8cm,width=0.7\textwidth]{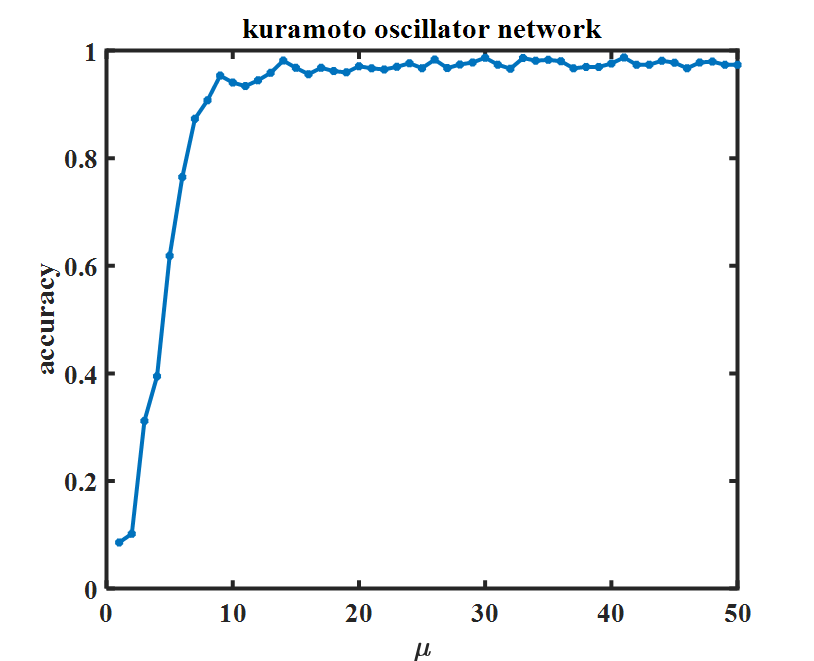}
	\caption{The inference accuracy for the Kuramoto network ($N=32$) with different natural frequency ranges.}
	\label{fig:fig13}
\end{figure}

If the network is fully synchronized, all nodes behave in essentially the same way. The time series look similar apart from a phase or a scaling factor, so it is hard to carry out the inference (see Supplementary Fig.~\ref{fig:FHNsyn}). Generally speaking, the more synchronized the network, the harder it is to infer the structure or coupling. Taking the Kuramoto oscillator network as an example, the natural frequency $\omega$ of each
node follows the uniform distribution $W(0,\mu)$, and a larger $\mu$ indicates that the dynamics has higher heterogeneity. As shown in the Fig. ~\ref{fig:fig13}, for $\mu\in [1,10]$, with the increase of $\mu$, the inference accuracy increases significantly. When $\mu$ reaches $10$, the accuracy is over $95\%$, and fluctuates around this value upon further increment of $\mu$.
%%%%%%%%%%%%%%%%%%%%%%%%%%%%%%%%%%%
%%%%%%%%%%%%%%%%%%%%%%%%%%%%%%%%%%%%%%%%%%%%

\begin{figure}[htbp]
	\centering
	\includegraphics[height=8cm,width=0.85\textwidth]{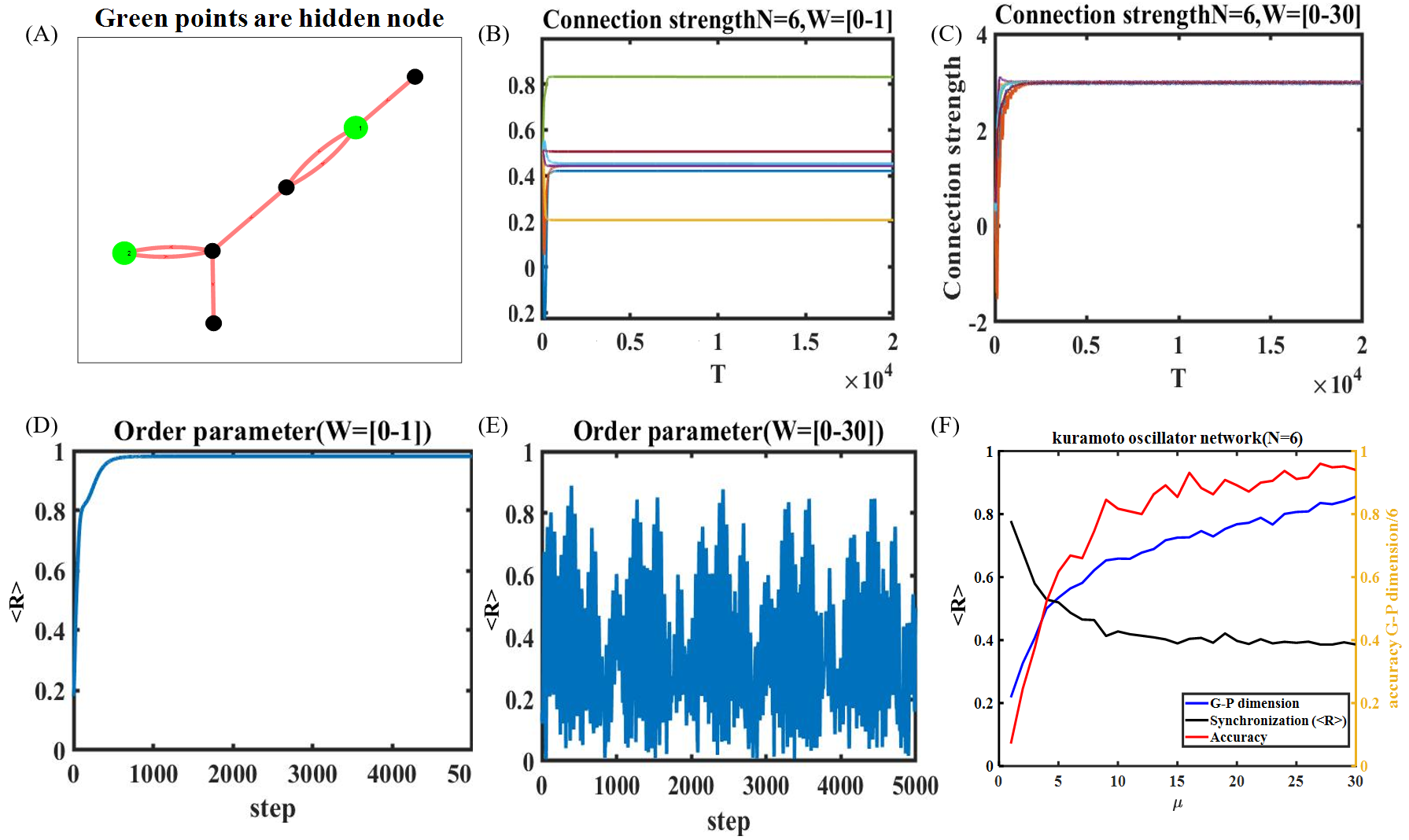}
	\caption{Reconstruction {\em vs} synchronization. (A) A network of six Kuramoto oscillators with coupling strength $K=3$, where the green dots are hidden nodes. (B) The inference of the coupling strengths at $\mu=1$. (C) The inference of the coupling strengths at $\mu=30$. (D) The convergence of the order parameter $<R>\to 1$ at $\mu=1$.(E) The oscillatory behaviour of the order parameter at $\mu=30$. (F) The dependence on $\mu$ of the G-P dimension (normalization by dividing with the maximum value $6$), the order parameter $<R>$ and the reconstruction accuracy.}
	\label{fig:fig14}
\end{figure}

In order to address in more detail the effect of synchronization on inference, we study a network of six oscillators coupled in a specific configuration as shown in Fig.~\ref{fig:fig14}(A), two of which are assumed to be hidden nodes. The natural frequencies of the oscillators conform to a uniform distribution in $[0,1]$ and the coupling strength is set to $3$. In Fig.~\ref{fig:fig14}(B), the coupling strengths indeed converge, but to completely wrong values, since in this case the oscillators fully synchronize. Nevertheless, if the distribution of the natural frequencies is expanded to a larger interval $[0,30]$ so that synchronization is not reached, our algorithm is able to dig out the coupling strengths as shown in Fig.~\ref{fig:fig14}(C). To quantify the performance of the algorithm at different levels of synchronization, we employ two statistical indicators - the order parameter $0\leq R \leq 1$ defined in Eq.~(\ref{eq:order}) and the G-P dimension~\cite{1983Characterization} for the oscillator network. The order parameter reflects the degree of synchronization, which is between $0$ (fully random) and $1$ (fully synchronous).  The fractal dimension gives the amount of entropy contained in the time series. The larger the value, the more entropy the time series contains. 

When the natural frequency is distributed in $[0,1]$,  $<R>=0.97$ as given in Fig.~\ref{fig:fig14}(D), which is close to a full synchronization. Accordingly, the G-P dimension is only $1.45$, indicating that the time series contains very little variability besides the nearly periodic oscillation, which makes it hard to distinguish different nodes. However, the entropy contained in the time series considerably increases if the frequency distribution is extended to the interval $[0,30]$, since the order parameter and the G-P dimension turn $0.38$ and $5.2$ respectively as plotted in Fig.~\ref{fig:fig14}(E), signalling little synchronization in the network. To see clearly the influence of heterogeneity induced by the frequency distribution $[0,\mu]$,  we scan different values of $\mu$ and depict the results in Fig.~\ref{fig:fig14}(F). The order parameter of the system decreases while the normalized dimensionality gradually increases, which indicates the waning of the synchronization. As expected, the accuracy of the reconstruction is gradually increasing and approaching $100\%$. 

\section*{Discussion}
Reconstructing system dynamics and interaction topology is an important yet challenging problem, especially with only partial information of the system state and in the presence of noise. In this article, we proposed a new framework for the reconstruction based on minimization of a carefully designed cost functional. To proceed in the gradient-descent direction in a continuous manner, a new system of differential equations are derived which serve to drive both the unknown observables and parameters to the correct values with the help of given time series. Several examples including the coupled Lorenz system, the FHN neural networks, the coupled Kuramoto oscillators and a gene regulation network (Supplementary Fig.~\ref{fig:gene}) are used to demonstrate validity of the current technique. Because of the evolution character, the method could be conveniently utilized to reconstruct complex dynamics and equations with nonlinear parameter dependence, or to monitor system reconfiguration changes on the scene.  
\par
With small noise in the system, the method is still useful while a direct application may fail in the presence of strong noise. So, it would be very interesting and helpful to extend the current frame to treat systems contaminated with strong noise. One possible way is to suppress noise in the available data with various existing techniques, for example, by using Wang's method~\cite{2021Reconstruction} to give a clean orbit first. Nevertheless, strong noise could even move the average orbit, which may be regarded as a renormalizing effect originating from the ignored degrees of freedom. It is desirable to utilize the framework here to renormalize complex dynamics with only slow degrees of freedom that are of our interest, treating those fast degrees as noise. 
\par
For small networks, a full connection assumption is made during the reconstruction and then the true network structure and connection strength are conveniently deduced with our method. But for large networks, prior knowledge of network topology is still needed since the number of connections grows quadratically with the network size while the available information increases only linearly. In practice, most networks are sparse and various techniques exist to infer possible connections. Even if the inferred links are spurious due to indirect interaction or confounding effects, our method still works. Nevertheless, it should be of great practical interest to extend the current framework to directly treat large-scale reconstruction without resorting to other inference programs beforehand. 
\par 
In the current work, the form of equations of motion is assumed known and only unknown parameters need to be deduced. Compared to existing algorithms~\cite{2017Brunton,Kera2016NoisetolerantAM,2022Autonomousinference,2001Structure}, this is not a serious restriction since we can always extend the expansion basis on the right hand side of the equation to include possible new terms. The cost is that more parameters should be deduced during the reconstruction, which is not a problem if the number of parameters is not over large. Of course, it remains a big challenge to deduce the underlying laws of motion purely from observation without much prior assumption although some progress has been made with machine learning~\cite{Cuomo2022ScientificML,Udrescu2019AIFA,PhysRevE106034315,RAISSI2019686}. It would be very interesting if we could combine the current framework with these cutting-edge techniques.  

\par
If the given information is too little, it seems impossible to deduce much as expected in general. Synchronization effectively reduces the fractal dimensions of the time series and makes it hard to carry out reconstruction, which also brings a lot of interesting problems. To get out of synchronization, we may consider transient dynamics or supply noise to unfold the reduced dimension. If it is impossible to apply external perturbation, it may be interesting to design an effective model to describe the reduced dynamics as indicated above. If synchronization is partial as in most cases, it is essential to determine and measure typical nodes to enable an equivalent modelling. In this case, some nodes may contain more information than others, how to find these is also an intriguing problem.
\par
In all, from the given information, how much we can deduce about the dynamics and structure of a system is an interesting problem both in theory and in application. We hope that the presented framework in the current paper will provide an alternative route for further investigation.

\section*{Conflict of interest}
The authors declare that they have no conflict of interest.

\bibliography{sample}

\begin{thebibliography}{10}
\urlstyle{rm}
\expandafter\ifx\csname url\endcsname\relax
  \def\url#1{\texttt{#1}}\fi
\expandafter\ifx\csname urlprefix\endcsname\relax\def\urlprefix{URL }\fi
\expandafter\ifx\csname doiprefix\endcsname\relax\def\doiprefix{DOI: }\fi
\providecommand{\bibinfo}[2]{#2}
\providecommand{\eprint}[2][]{\url{#2}}

\bibitem{1Sch1992Nonlinear}
\bibinfo{author}{Schütte, R.} \& \bibinfo{author}{Zelewski, S.}
\newblock \emph{\bibinfo{title}{Nonlinear Modeling and Forecasting}}
  (\bibinfo{publisher}{Addison-Wesley Publishing Company},
  \bibinfo{year}{1992}).

\bibitem{2Winkel1995Application}
\bibinfo{author}{Winkel, P.}
\newblock \bibinfo{journal}{\bibinfo{title}{Application of time series analysis
  in the clinical setting}}.
\newblock {\emph{\JournalTitle{Scand J Clin Lab Inv}}}
  \textbf{\bibinfo{volume}{55}}, \bibinfo{pages}{11--16}
  (\bibinfo{year}{1995}).

\bibitem{3pasten2018time}
\bibinfo{author}{Past{\'e}n, D.}, \bibinfo{author}{Czechowski, Z.} \&
  \bibinfo{author}{Toledo, B.}
\newblock \bibinfo{journal}{\bibinfo{title}{Time series analysis in earthquake
  complex networks}}.
\newblock {\emph{\JournalTitle{Chaos}}} \textbf{\bibinfo{volume}{28}},
  \bibinfo{pages}{083128} (\bibinfo{year}{2018}).

\bibitem{4Gouveia2000Time}
\bibinfo{author}{Gouveia, N.}
\newblock \bibinfo{journal}{\bibinfo{title}{Time series analysis of air
  pollution and mortality: effects by cause, age and socioeconomic status}}.
\newblock {\emph{\JournalTitle{J Epidemiol Commun H}}}
  \textbf{\bibinfo{volume}{54}}, \bibinfo{pages}{750--755}
  (\bibinfo{year}{2000}).

\bibitem{5caldarelli2013reconstructing}
\bibinfo{author}{Caldarelli, G.}, \bibinfo{author}{Chessa, A.},
  \bibinfo{author}{Pammolli, F.}, \bibinfo{author}{Gabrielli, A.} \&
  \bibinfo{author}{Puliga, M.}
\newblock \bibinfo{journal}{\bibinfo{title}{Reconstructing a credit network}}.
\newblock {\emph{\JournalTitle{Nat Phys}}} \textbf{\bibinfo{volume}{9}},
  \bibinfo{pages}{125--126} (\bibinfo{year}{2013}).

\bibitem{62009The}
\bibinfo{author}{Donges, J.~F.}, \bibinfo{author}{Zou, Y.},
  \bibinfo{author}{Marwan, N.} \& \bibinfo{author}{Kurths, J.}
\newblock \bibinfo{journal}{\bibinfo{title}{The backbone of the climate
  network}}.
\newblock {\emph{\JournalTitle{Europhys Lett}}} \textbf{\bibinfo{volume}{87}},
  \bibinfo{pages}{0295--5075} (\bibinfo{year}{2009}).

\bibitem{7chen2014the}
\bibinfo{author}{Chen, G.~R.}
\newblock \bibinfo{journal}{\bibinfo{title}{The china power grid: a network
  science perspective}}.
\newblock {\emph{\JournalTitle{Natl Sci Rev}}} \textbf{\bibinfo{volume}{1}},
  \bibinfo{pages}{368} (\bibinfo{year}{2014}).

\bibitem{81998Genome}
\bibinfo{author}{The, C.} \& \bibinfo{author}{Consortium, E.~S.}
\newblock \bibinfo{journal}{\bibinfo{title}{Genome sequence of the nematode c.
  elegans: A platform for investigating biology}}.
\newblock {\emph{\JournalTitle{Science}}} \textbf{\bibinfo{volume}{282}},
  \bibinfo{pages}{2012--2018} (\bibinfo{year}{1998}).

\bibitem{92003Inferring}
\bibinfo{author}{Gardner, T.~S.}, \bibinfo{author}{Bernardo, D.~D.},
  \bibinfo{author}{Lorenz, D.} \& \bibinfo{author}{Collins, J.~J.}
\newblock \bibinfo{journal}{\bibinfo{title}{Inferring genetic networks and
  identifying compound mode of action via expression profiling}}.
\newblock {\emph{\JournalTitle{Science}}} \textbf{\bibinfo{volume}{301}},
  \bibinfo{pages}{102--105} (\bibinfo{year}{2003}).

\bibitem{102006algorithm}
\bibinfo{author}{Margolin, A.}, \bibinfo{author}{Nemenman, I.} \&
  \bibinfo{author}{Basso, K.}
\newblock \bibinfo{journal}{\bibinfo{title}{An algorithm for the reconstruction
  of gene regulatory networks in a mammalian cellular context}}.
\newblock {\emph{\JournalTitle{BMC Bioinform}}} \textbf{\bibinfo{volume}{1}},
  \bibinfo{pages}{s1--s7} (\bibinfo{year}{2006}).

\bibitem{2009Inferring}
\bibinfo{author}{Eagle, N.}, \bibinfo{author}{Pentland, A.} \&
  \bibinfo{author}{Lazer, D.}
\newblock \bibinfo{journal}{\bibinfo{title}{Inferring friendship network
  structure by using mobile phone data}}.
\newblock {\emph{\JournalTitle{Pro Natl Acad Sci USA}}}
  \textbf{\bibinfo{volume}{106}} (\bibinfo{year}{2009}).

\bibitem{2012Model}
\bibinfo{author}{Stetter, O.}, \bibinfo{author}{Battaglia, D.},
  \bibinfo{author}{Soriano, J.} \& \bibinfo{author}{Geisel, T.}
\newblock \bibinfo{journal}{\bibinfo{title}{Model-free reconstruction of
  excitatory neuronal connectivity from calcium imaging signals}}.
\newblock {\emph{\JournalTitle{Plos Comput Biol}}}
  \textbf{\bibinfo{volume}{8}}, \bibinfo{pages}{1002653}
  (\bibinfo{year}{2012}).

\bibitem{B2015Systems}
\bibinfo{author}{Palsson, B.}
\newblock \emph{\bibinfo{title}{Systems Biology- Properties of Reconstructed
  Networks}} (\bibinfo{publisher}{Cambridge University Press},
  \bibinfo{year}{2015}).

\bibitem{2006Complex}
\bibinfo{author}{Boccaletti, S.}, \bibinfo{author}{Latora, V.},
  \bibinfo{author}{Moreno, Y.}, \bibinfo{author}{Chavez, M.} \&
  \bibinfo{author}{Hwang, D.~U.}
\newblock \bibinfo{journal}{\bibinfo{title}{Complex networks: Structure and
  dynamics}}.
\newblock {\emph{\JournalTitle{Complex Syst and Complexity Sci}}}
  \textbf{\bibinfo{volume}{424}}, \bibinfo{pages}{175–308}
  (\bibinfo{year}{2006}).

\bibitem{2004Forecast}
\bibinfo{author}{Hufnagel, L.}, \bibinfo{author}{Brockmann, D.} \&
  \bibinfo{author}{Geisel, T.}
\newblock \bibinfo{journal}{\bibinfo{title}{Forecast and control of epidemics
  in a globalized world}}.
\newblock {\emph{\JournalTitle{Pro Natl Acad Sci USA}}}
  \textbf{\bibinfo{volume}{101}}, \bibinfo{pages}{15124--15129}
  (\bibinfo{year}{2004}).

\bibitem{2009Complex}
\bibinfo{author}{Bullmore, E.} \& \bibinfo{author}{Sporns, O.}
\newblock \bibinfo{journal}{\bibinfo{title}{Complex brain networks: Graph
  theoretical analysis of structural and functional systems}}.
\newblock {\emph{\JournalTitle{Nat Rev Neurosci}}}
  \textbf{\bibinfo{volume}{10}}, \bibinfo{pages}{186--198}
  (\bibinfo{year}{2009}).

\bibitem{Mcharakspar2017Dynamic}
\bibinfo{author}{Breakspear, M.}
\newblock \bibinfo{journal}{\bibinfo{title}{Dynamic models of large-scale brain
  activity}}.
\newblock {\emph{\JournalTitle{Nat Neurosci}}} \textbf{\bibinfo{volume}{20}},
  \bibinfo{pages}{340--352} (\bibinfo{year}{2017}).

\bibitem{2003Newman}
\bibinfo{author}{Newman, M.}
\newblock \bibinfo{journal}{\bibinfo{title}{The structure and function of
  complex networks.}}
\newblock {\emph{\JournalTitle{SIAM Review}}} \textbf{\bibinfo{volume}{45}}
  (\bibinfo{year}{2003}).

\bibitem{33PhysRevLett}
\bibinfo{author}{Timme, M.}
\newblock \bibinfo{journal}{\bibinfo{title}{Revealing network connectivity from
  response dynamics}}.
\newblock {\emph{\JournalTitle{Phys Rev Lett}}} \textbf{\bibinfo{volume}{98}},
  \bibinfo{pages}{224101} (\bibinfo{year}{2007}).

\bibitem{34shandilya2011inferring}
\bibinfo{author}{Shandilya, S.~G.} \& \bibinfo{author}{Timme, M.}
\newblock \bibinfo{journal}{\bibinfo{title}{Inferring network topology from
  complex dynamics}}.
\newblock {\emph{\JournalTitle{New J Phys}}} \textbf{\bibinfo{volume}{13}},
  \bibinfo{pages}{87--92} (\bibinfo{year}{2011}).

\bibitem{352008Reconstructing}
\bibinfo{author}{Napoletani, D.} \& \bibinfo{author}{Sauer, T.~D.}
\newblock \bibinfo{journal}{\bibinfo{title}{Reconstructing the topology of
  sparsely connected dynamical networks}}.
\newblock {\emph{\JournalTitle{Phys Rev E}}} \textbf{\bibinfo{volume}{77}},
  \bibinfo{pages}{026103} (\bibinfo{year}{2008}).

\bibitem{2016Discovering}
\bibinfo{author}{Steven, L.} \emph{et~al.}
\newblock \bibinfo{journal}{\bibinfo{title}{Discovering governing equations
  from data by sparse identification of nonlinear dynamical systems}}.
\newblock {\emph{\JournalTitle{Pro Natl Acad Sci USA}}}
  \textbf{\bibinfo{volume}{113}}, \bibinfo{pages}{3932–3937}
  (\bibinfo{year}{2016}).

\bibitem{2017Model}
\bibinfo{author}{Casadiego, J.}, \bibinfo{author}{Nitzan, M.},
  \bibinfo{author}{Hallerberg, S.} \& \bibinfo{author}{Timme, M.}
\newblock \bibinfo{journal}{\bibinfo{title}{Model-free inference of direct
  network interactions from nonlinear collective dynamics}}.
\newblock {\emph{\JournalTitle{Nat Commun}}} \textbf{\bibinfo{volume}{8}},
  \bibinfo{pages}{1–10} (\bibinfo{year}{2017}).

\bibitem{Lorenz1991DimensionOW}
\bibinfo{author}{Lorenz, E.~N.}
\newblock \bibinfo{journal}{\bibinfo{title}{Dimension of weather and climate
  attractors}}.
\newblock {\emph{\JournalTitle{Nature}}} \textbf{\bibinfo{volume}{353}},
  \bibinfo{pages}{241--244} (\bibinfo{year}{1991}).

\bibitem{112007Partial}
\bibinfo{author}{Frenzel, S.} \& \bibinfo{author}{Pompe, B.}
\newblock \bibinfo{journal}{\bibinfo{title}{Partial mutual information for
  coupling analysis of multivariate time series}}.
\newblock {\emph{\JournalTitle{Phys Rev Lett}}} \textbf{\bibinfo{volume}{99}},
  \bibinfo{pages}{204101} (\bibinfo{year}{2007}).

\bibitem{122011Momentary}
\bibinfo{author}{Pompe, B.} \& \bibinfo{author}{Runge, J.}
\newblock \bibinfo{journal}{\bibinfo{title}{Momentary information transfer as a
  coupling measure of time series}}.
\newblock {\emph{\JournalTitle{Phys Rev E}}} \textbf{\bibinfo{volume}{83}},
  \bibinfo{pages}{051122} (\bibinfo{year}{2011}).

\bibitem{Sch2000}
\bibinfo{author}{Schreiber, T.}
\newblock \bibinfo{journal}{\bibinfo{title}{Measuring information transfer}}.
\newblock {\emph{\JournalTitle{Phys Rev Lett}}} \textbf{\bibinfo{volume}{85}},
  \bibinfo{pages}{461--464} (\bibinfo{year}{2000}).

\bibitem{522008Kernel}
\bibinfo{author}{Marinazzo, D.}, \bibinfo{author}{Pellicoro, M.} \&
  \bibinfo{author}{Stramaglia, S.}
\newblock \bibinfo{journal}{\bibinfo{title}{Kernel method for nonlinear granger
  causality}}.
\newblock {\emph{\JournalTitle{Phys Rev Lett}}} \textbf{\bibinfo{volume}{100}},
  \bibinfo{pages}{144103} (\bibinfo{year}{2008}).

\bibitem{1992Fitting}
\bibinfo{author}{Baake, E.}, \bibinfo{author}{Baake, M.},
  \bibinfo{author}{Bock, H.~G.} \& \bibinfo{author}{Briggs, K.~M.}
\newblock \bibinfo{journal}{\bibinfo{title}{Fitting ordinary differential
  equations to chaotic data}}.
\newblock {\emph{\JournalTitle{Physical Review A}}}
  \textbf{\bibinfo{volume}{45}}, \bibinfo{pages}{5524--5529}
  (\bibinfo{year}{1992}).

\bibitem{2017Brunton}
\bibinfo{author}{Rudy, S.~H.}, \bibinfo{author}{Brunton, S.~L.},
  \bibinfo{author}{Proctor, J.~L.} \& \bibinfo{author}{Kutz, J.~N.}
\newblock \bibinfo{journal}{\bibinfo{title}{Data-driven discovery of partial
  differential equations}}.
\newblock {\emph{\JournalTitle{Science advances}}}
  \textbf{\bibinfo{volume}{3}}, \bibinfo{pages}{e1602614}
  (\bibinfo{year}{2017}).

\bibitem{13yu2006estimating}
\bibinfo{author}{Yu, D.}, \bibinfo{author}{Righero, M.} \&
  \bibinfo{author}{Kocarev, L.}
\newblock \bibinfo{journal}{\bibinfo{title}{Estimating topology of networks}}.
\newblock {\emph{\JournalTitle{Phys Rev Lett}}} \textbf{\bibinfo{volume}{97}},
  \bibinfo{pages}{188701} (\bibinfo{year}{2006}).

\bibitem{142007Topology}
\bibinfo{author}{Zhou, J.} \& \bibinfo{author}{Lu, J.~A.}
\newblock \bibinfo{journal}{\bibinfo{title}{Topology identification of weighted
  complex dynamical networks}}.
\newblock {\emph{\JournalTitle{Physica A}}} \textbf{\bibinfo{volume}{386}},
  \bibinfo{pages}{481--491} (\bibinfo{year}{2007}).

\bibitem{152008Synchronization}
\bibinfo{author}{Wu, X.}
\newblock \bibinfo{journal}{\bibinfo{title}{Synchronization-based topology
  identification of weighted general complex dynamical networks with
  time-varying coupling delay}}.
\newblock {\emph{\JournalTitle{Physica A}}} \textbf{\bibinfo{volume}{387}},
  \bibinfo{pages}{997--1008} (\bibinfo{year}{2008}).

\bibitem{162009Structure}
\bibinfo{author}{Liu, H.}, \bibinfo{author}{Lu, J.~A.}, \bibinfo{author}{Lü,
  J.} \& \bibinfo{author}{Hill, D.~J.}
\newblock \bibinfo{journal}{\bibinfo{title}{Structure identification of
  uncertain general complex dynamical networks with time delay}}.
\newblock {\emph{\JournalTitle{Automatica}}} \textbf{\bibinfo{volume}{45}},
  \bibinfo{pages}{1799--1807} (\bibinfo{year}{2009}).

\bibitem{17Parlitz1996Estimating}
\bibinfo{author}{U.Parlitz}.
\newblock \bibinfo{journal}{\bibinfo{title}{Estimating model parameters from
  time series by autosynchronization}}.
\newblock {\emph{\JournalTitle{Phys Rev Lett}}} \textbf{\bibinfo{volume}{76}},
  \bibinfo{pages}{1232--1235} (\bibinfo{year}{1996}).

\bibitem{182007Estimating}
\bibinfo{author}{Tao, C.}, \bibinfo{author}{Zhang, Y.} \&
  \bibinfo{author}{Jiang, J.~J.}
\newblock \bibinfo{journal}{\bibinfo{title}{Estimating system parameters from
  chaotic time series with synchronization optimized by a genetic algorithm}}.
\newblock {\emph{\JournalTitle{Phys Rev E}}} \textbf{\bibinfo{volume}{76}},
  \bibinfo{pages}{016209} (\bibinfo{year}{2007}).

\bibitem{43Zoran2011Network}
\bibinfo{author}{Levnaji, Z.} \& \bibinfo{author}{Pikovsky, A.}
\newblock \bibinfo{journal}{\bibinfo{title}{Network reconstruction from random
  phase resetting}}.
\newblock {\emph{\JournalTitle{Phys Rev Lett}}} \textbf{\bibinfo{volume}{107}},
  \bibinfo{pages}{34101--34101} (\bibinfo{year}{2011}).

\bibitem{19ren2010noise}
\bibinfo{author}{Ren, J.}, \bibinfo{author}{Wang, W.~X.}, \bibinfo{author}{Li,
  B.} \& \bibinfo{author}{Lai, Y.~C.}
\newblock \bibinfo{journal}{\bibinfo{title}{Noise bridges dynamical correlation
  and topology in coupled oscillator networks}}.
\newblock {\emph{\JournalTitle{Phys Rev Lett}}} \textbf{\bibinfo{volume}{104}},
  \bibinfo{pages}{058701} (\bibinfo{year}{2010}).

\bibitem{20Emily2014Erratum}
\bibinfo{author}{Ching, E.}, \bibinfo{author}{Lai, P.~Y.} \&
  \bibinfo{author}{Leung, C.~Y.}
\newblock \bibinfo{journal}{\bibinfo{title}{Extracting connectivity from
  dynamics of networks with uniform bidirectional coupling}}.
\newblock {\emph{\JournalTitle{Phys Rev E}}} \textbf{\bibinfo{volume}{88}},
  \bibinfo{pages}{042817} (\bibinfo{year}{2013}).

\bibitem{21ching2017reconstructing}
\bibinfo{author}{Ching, E.} \& \bibinfo{author}{Tam, H.~C.}
\newblock \bibinfo{journal}{\bibinfo{title}{Reconstructing links in directed
  networks from noisy dynamics}}.
\newblock {\emph{\JournalTitle{Phys Rev E}}} \textbf{\bibinfo{volume}{95}},
  \bibinfo{pages}{010301} (\bibinfo{year}{2017}).

\bibitem{22zhang2015solving}
\bibinfo{author}{Zhang, Z.} \emph{et~al.}
\newblock \bibinfo{journal}{\bibinfo{title}{Solving the inverse problem of
  noise-driven dynamic networks}}.
\newblock {\emph{\JournalTitle{Phys Rev E}}} \textbf{\bibinfo{volume}{91}},
  \bibinfo{pages}{012814} (\bibinfo{year}{2015}).

\bibitem{232017Reconstruction}
\bibinfo{author}{Yang, C.}, \bibinfo{author}{Zhang, Z.}, \bibinfo{author}{Chen,
  T.}, \bibinfo{author}{Wang, S.} \& \bibinfo{author}{Hu, G.}
\newblock \bibinfo{journal}{\bibinfo{title}{Reconstruction of noise-driven
  nonlinear networks from node outputs by using high-order correlations}}.
\newblock {\emph{\JournalTitle{Sci. Rep}}} \textbf{\bibinfo{volume}{7}},
  \bibinfo{pages}{44639} (\bibinfo{year}{2017}).

\bibitem{2021Reconstruction}
\bibinfo{author}{Wang, J.}, \bibinfo{author}{Yan, Z.}, \bibinfo{author}{Gui,
  L.}, \bibinfo{author}{Xu, K.} \& \bibinfo{author}{Lan, Y.}
\newblock \bibinfo{journal}{\bibinfo{title}{Reconstruction of nonlinear flows
  from noisy time series}}.
\newblock {\emph{\JournalTitle{Nonliner Dynam}}}
  \textbf{\bibinfo{volume}{108}}, \bibinfo{pages}{3887–--3902}
  (\bibinfo{year}{2022}).

\bibitem{2022Autonomousinference}
\bibinfo{author}{Gao, T.-T.} \& \bibinfo{author}{Yan, G.}
\newblock \bibinfo{journal}{\bibinfo{title}{Autonomous inference of complex
  network dynamics from incomplete and noisy data}}.
\newblock {\emph{\JournalTitle{Nat Comput Sci}}} \textbf{\bibinfo{volume}{2}},
  \bibinfo{pages}{160--168} (\bibinfo{year}{2022}).

\bibitem{24201990Reconstructing}
\bibinfo{author}{Breeden, J.~L.} \& \bibinfo{author}{Hübler, A.}
\newblock \bibinfo{journal}{\bibinfo{title}{Reconstructing equations of motion
  from experimental data with unobserved variables}}.
\newblock {\emph{\JournalTitle{Phys Rev A}}} \textbf{\bibinfo{volume}{42}},
  \bibinfo{pages}{5817} (\bibinfo{year}{1990}).

\bibitem{251987Construction}
\bibinfo{author}{Cremers, J.} \& \bibinfo{author}{Hübler, A.}
\newblock \bibinfo{journal}{\bibinfo{title}{Construction of differential
  equations from experimental data}}.
\newblock {\emph{\JournalTitle{Z Naturforsch A}}}
  \textbf{\bibinfo{volume}{42}}, \bibinfo{pages}{797--802}
  (\bibinfo{year}{1987}).

\bibitem{26James1987Equations}
\bibinfo{author}{James, P.}, \bibinfo{author}{Crutchfield, B.} \&
  \bibinfo{author}{McNamara, S.}
\newblock \bibinfo{journal}{\bibinfo{title}{Equations of motion from a data
  series}}.
\newblock {\emph{\JournalTitle{Complex Syst}}} \textbf{\bibinfo{volume}{1}}
  (\bibinfo{year}{1987}).

\bibitem{27gouesbet1991reconstruction}
\bibinfo{author}{Gouesbet, G.}
\newblock \bibinfo{journal}{\bibinfo{title}{Reconstruction of standard and
  inverse vector fields equivalent to a rossler system}}.
\newblock {\emph{\JournalTitle{Phys Rev A}}} \textbf{\bibinfo{volume}{44}},
  \bibinfo{pages}{6264--6280} (\bibinfo{year}{1991}).

\bibitem{28gouesbet1992reconstruction}
\bibinfo{author}{Gouesbet, G.}
\newblock \bibinfo{journal}{\bibinfo{title}{Reconstruction of vector fields:
  The case of the lorenz system}}.
\newblock {\emph{\JournalTitle{Phys Rev A}}} \textbf{\bibinfo{volume}{46}},
  \bibinfo{pages}{1784--1796} (\bibinfo{year}{1992}).

\bibitem{2011Network}
\bibinfo{author}{Wang, W.~X.}, \bibinfo{author}{Lai, Y.~C.},
  \bibinfo{author}{Grebogi, C.} \& \bibinfo{author}{Ye, J.}
\newblock \bibinfo{journal}{\bibinfo{title}{Network reconstruction based on
  evolutionary-game data via compressive sensing}}.
\newblock {\emph{\JournalTitle{Phys Rev X}}} \textbf{\bibinfo{volume}{1}},
  \bibinfo{pages}{021021} (\bibinfo{year}{2011}).

\bibitem{292012Detecting}
\bibinfo{author}{Su, R.~Q.}, \bibinfo{author}{Wang, W.~X.} \&
  \bibinfo{author}{Lai, Y.~C.}
\newblock \bibinfo{journal}{\bibinfo{title}{Detecting hidden nodes in complex
  networks from time series}}.
\newblock {\emph{\JournalTitle{Phys Rev E}}} \textbf{\bibinfo{volume}{85}},
  \bibinfo{pages}{1149--1164} (\bibinfo{year}{2012}).

\bibitem{302014Uncovering}
\bibinfo{author}{Su, R.~Q.}, \bibinfo{author}{Lai, Y.~C.},
  \bibinfo{author}{Wang, W.~X.} \& \bibinfo{author}{Do, Y.}
\newblock \bibinfo{journal}{\bibinfo{title}{Uncovering hidden nodes in complex
  networks in the presence of noise}}.
\newblock {\emph{\JournalTitle{Sci. Rep}}} \textbf{\bibinfo{volume}{4}}
  (\bibinfo{year}{2014}).

\bibitem{31Ri2016Data}
\bibinfo{author}{Su, R.~Q.}, \bibinfo{author}{Wang, W.~X.},
  \bibinfo{author}{Wang, X.} \& \bibinfo{author}{Lai, Y.~C.}
\newblock \bibinfo{journal}{\bibinfo{title}{Data-based reconstruction of
  complex geospatial networks, nodal positioning and detection of hidden
  nodes}}.
\newblock {\emph{\JournalTitle{Roy Soc Open Sci}}} \textbf{\bibinfo{volume}{3}}
  (\bibinfo{year}{2016}).

\bibitem{322014Reconstructing}
\bibinfo{author}{Shen, Z.}, \bibinfo{author}{Wang, W.~X.},
  \bibinfo{author}{Fan, Y.}, \bibinfo{author}{Di, Z.} \& \bibinfo{author}{Lai,
  Y.~C.}
\newblock \bibinfo{journal}{\bibinfo{title}{Reconstructing propagation networks
  with natural diversity and identifying hidden sources}}.
\newblock {\emph{\JournalTitle{Nat Commun}}} \textbf{\bibinfo{volume}{5}},
  \bibinfo{pages}{4323} (\bibinfo{year}{2014}).

\bibitem{36Xiaoqun2012Inferring}
\bibinfo{author}{Xiaoqun, W.}, \bibinfo{author}{Weihan, W.} \&
  \bibinfo{author}{WeiXing, Z.}
\newblock \bibinfo{journal}{\bibinfo{title}{Inferring topologies of complex
  networks with hidden variables}}.
\newblock {\emph{\JournalTitle{Phys Rev E}}} \textbf{\bibinfo{volume}{86}}
  (\bibinfo{year}{2012}).

\bibitem{37GUO200879}
\bibinfo{author}{Guo, S.}, \bibinfo{author}{Seth, A.~K.},
  \bibinfo{author}{Kendrick, K.~M.}, \bibinfo{author}{Zhou, C.} \&
  \bibinfo{author}{Feng, J.}
\newblock \bibinfo{journal}{\bibinfo{title}{Partial granger
  causality—eliminating exogenous inputs and latent variables}}.
\newblock {\emph{\JournalTitle{J Neurosci Meth}}}
  \textbf{\bibinfo{volume}{172}}, \bibinfo{pages}{79--93}
  (\bibinfo{year}{2008}).

\bibitem{38chen2017reconstruction}
\bibinfo{author}{Yang, C.}, \bibinfo{author}{Zhang, Z.}, \bibinfo{author}{Chen,
  T.~Y.}, \bibinfo{author}{Wang, S.~H.} \& \bibinfo{author}{Gang, H.}
\newblock \bibinfo{journal}{\bibinfo{title}{Reconstruction of noise-driven
  nonlinear dynamic networks with some hidden nodes}}.
\newblock {\emph{\JournalTitle{Sci China Phys Mech}}}
  \textbf{\bibinfo{volume}{60}}, \bibinfo{pages}{46} (\bibinfo{year}{2017}).

\bibitem{39zhang2017network}
\bibinfo{author}{Zhang, Z.}, \bibinfo{author}{Yang, C.} \&
  \bibinfo{author}{Gang, H.}
\newblock \bibinfo{journal}{\bibinfo{title}{Network reconstructions with
  partially available data}}.
\newblock {\emph{\JournalTitle{Front Phys China}}}
  \textbf{\bibinfo{volume}{12}}, \bibinfo{pages}{117--123}
  (\bibinfo{year}{2017}).

\bibitem{41Wangsh}
\bibinfo{author}{Shi, R.}, \bibinfo{author}{Jiang, W.} \&
  \bibinfo{author}{Wang, S.}
\newblock \bibinfo{journal}{\bibinfo{title}{Detecting network structures from
  measurable data produced by dynamics with hidden variables}}.
\newblock {\emph{\JournalTitle{chaos}}} \textbf{\bibinfo{volume}{30}},
  \bibinfo{pages}{138} (\bibinfo{year}{2020}).

\bibitem{2013Extracting}
\bibinfo{author}{Ching, E.}, \bibinfo{author}{Lai, P.~Y.} \&
  \bibinfo{author}{Leung, C.~Y.}
\newblock \bibinfo{journal}{\bibinfo{title}{Extracting connectivity from
  dynamics of networks with uniform bidirectional coupling}}.
\newblock {\emph{\JournalTitle{Phys Rev E}}} \textbf{\bibinfo{volume}{88}},
  \bibinfo{pages}{042817} (\bibinfo{year}{2013}).

\bibitem{442018Effects}
\bibinfo{author}{Ching, E.} \& \bibinfo{author}{Tam, P.~H.}
\newblock \bibinfo{journal}{\bibinfo{title}{Effects of hidden nodes on the
  reconstruction of bidirectional networks}}.
\newblock {\emph{\JournalTitle{Phys Rev E}}} \textbf{\bibinfo{volume}{98}}
  (\bibinfo{year}{2018}).

\bibitem{45lorenz1963deterministic}
\bibinfo{author}{Lorenz, E.~N.}
\newblock \bibinfo{journal}{\bibinfo{title}{Deterministic nonperiodic flow}}.
\newblock {\emph{\JournalTitle{J Atmos Sci}}} \textbf{\bibinfo{volume}{20}},
  \bibinfo{pages}{130--141} (\bibinfo{year}{1963}).

\bibitem{ottChaos}
\bibinfo{author}{Ott, E.}
\newblock \emph{\bibinfo{title}{Chaos in Dynamical Systems}}
  (\bibinfo{publisher}{Cambridge University Press}, \bibinfo{year}{2002}).

\bibitem{Kuramoto}
\bibinfo{author}{Kuramoto, Y.}
\newblock \bibinfo{title}{Self-entrainment of a population of coupled
  non-linear oscillators}.
\newblock In \emph{\bibinfo{booktitle}{International Symposium on Mathematical
  Problems in Theoretical Physics}}, \bibinfo{pages}{420--422}
  (\bibinfo{publisher}{Springer Berlin Heidelberg}, \bibinfo{address}{Berlin,
  Heidelberg}, \bibinfo{year}{1975}).

\bibitem{PIETRAS20191}
\bibinfo{author}{Pietras, B.} \& \bibinfo{author}{Daffertshofer, A.}
\newblock \bibinfo{journal}{\bibinfo{title}{Network dynamics of coupled
  oscillators and phase reduction techniques}}.
\newblock {\emph{\JournalTitle{Physics Reports}}}
  \textbf{\bibinfo{volume}{819}}, \bibinfo{pages}{1--105}
  (\bibinfo{year}{2019}).

\bibitem{Acebrn2005TheKM}
\bibinfo{author}{Acebr{\'o}n, J.~A.}, \bibinfo{author}{Bonilla, L.~L.},
  \bibinfo{author}{Vicente, C. J.~P.}, \bibinfo{author}{Ritort, F.} \&
  \bibinfo{author}{Spigler, R.}
\newblock \bibinfo{journal}{\bibinfo{title}{The kuramoto model: A simple
  paradigm for synchronization phenomena}}.
\newblock {\emph{\JournalTitle{Reviews of Modern Physics}}}
  \textbf{\bibinfo{volume}{77}}, \bibinfo{pages}{137--185}
  (\bibinfo{year}{2005}).

\bibitem{1983Characterization}
\bibinfo{author}{Grassberger, P.} \& \bibinfo{author}{Procaccia, I.}
\newblock \bibinfo{journal}{\bibinfo{title}{Characterization of strange
  attractors}}.
\newblock {\emph{\JournalTitle{Phys Rev Lett}}} \textbf{\bibinfo{volume}{50}},
  \bibinfo{pages}{346} (\bibinfo{year}{1983}).

\bibitem{Kera2016NoisetolerantAM}
\bibinfo{author}{Kera, H.} \& \bibinfo{author}{Hasegawa, Y.}
\newblock \bibinfo{journal}{\bibinfo{title}{Noise-tolerant algebraic method for
  reconstruction of nonlinear dynamical systems}}.
\newblock {\emph{\JournalTitle{Nonlinear Dynamics}}}
  \textbf{\bibinfo{volume}{85}}, \bibinfo{pages}{675 -- 692}
  (\bibinfo{year}{2016}).

\bibitem{2001Structure}
\bibinfo{author}{Aguirre, L.}, \bibinfo{author}{Freitas, U.},
  \bibinfo{author}{Letellier, C.} \& \bibinfo{author}{Maquet, J.}
\newblock \bibinfo{journal}{\bibinfo{title}{Structure-selection techniques
  applied to continuous-time nonlinear models}}.
\newblock {\emph{\JournalTitle{Physica D}}} \textbf{\bibinfo{volume}{158}},
  \bibinfo{pages}{1--18} (\bibinfo{year}{2001}).

\bibitem{Cuomo2022ScientificML}
\bibinfo{author}{Cuomo, S.} \emph{et~al.}
\newblock \bibinfo{journal}{\bibinfo{title}{Scientific machine learning through
  physics–informed neural networks: Where we are and what’s next}}.
\newblock {\emph{\JournalTitle{Journal of Scientific Computing}}}
  \textbf{\bibinfo{volume}{92}} (\bibinfo{year}{2022}).

\bibitem{Udrescu2019AIFA}
\bibinfo{author}{Udrescu, S.-M.} \& \bibinfo{author}{Tegmark, M.}
\newblock \bibinfo{journal}{\bibinfo{title}{Ai feynman: A physics-inspired
  method for symbolic regression}}.
\newblock {\emph{\JournalTitle{Science Advances}}} \textbf{\bibinfo{volume}{6}}
  (\bibinfo{year}{2019}).

\bibitem{PhysRevE106034315}
\bibinfo{author}{Zhang, Y.} \emph{et~al.}
\newblock \bibinfo{journal}{\bibinfo{title}{Universal framework for
  reconstructing complex networks and node dynamics from discrete or continuous
  dynamics data}}.
\newblock {\emph{\JournalTitle{Phys. Rev. E}}} \textbf{\bibinfo{volume}{106}},
  \bibinfo{pages}{034315} (\bibinfo{year}{2022}).

\bibitem{RAISSI2019686}
\bibinfo{author}{Raissi, M.}, \bibinfo{author}{Perdikaris, P.} \&
  \bibinfo{author}{Karniadakis, G.}
\newblock \bibinfo{journal}{\bibinfo{title}{Physics-informed neural networks: A
  deep learning framework for solving forward and inverse problems involving
  nonlinear partial differential equations}}.
\newblock {\emph{\JournalTitle{Journal of Computational Physics}}}
  \textbf{\bibinfo{volume}{378}}, \bibinfo{pages}{686--707}
  (\bibinfo{year}{2019}).

\bibitem{2009Reconstructinggene}
\bibinfo{author}{Mazur, J.}, \bibinfo{author}{Ritter, D.},
  \bibinfo{author}{Reinelt, G.} \& \bibinfo{author}{Kaderali, L.}
\newblock \bibinfo{journal}{\bibinfo{title}{Reconstructing nonlinear dynamic
  models of gene regulation using stochastic sampling}}.
\newblock {\emph{\JournalTitle{BMC Bioinformatics}}}
  \textbf{\bibinfo{volume}{10}}, \bibinfo{pages}{448} (\bibinfo{year}{2009}).

\end{thebibliography}

\section*{Acknowledgements}
This work was supported by the Key Program of National Natural Science Foundation of China (No. 92067202), the National Natural Science Foundation of China under Grants No.11775035, the Fundamental Research Funds for the Central Universities from China (Grant ZDYY202102-1), and in part by the Fund of State Key Laboratory of Information Photonics and Optical Communications (Beijing University of Posts and Telecommunications) of China under Grant IPOC2021ZR02.

\section*{Author contributions statement}
Yueheng Lan, Lili Gui and Kun Xu conceived the research, Yueheng Lan and Zishuo Yan designed the research, Zishuo Yan performed the research, Zishuo Yan and Yueheng Lan analyzed the results, Zishuo Yan and Yueheng Lan wrote this manuscript. All authors reviewed the manuscript. 

\section*{Data availability}
The data that support the findings of this study are available from the corresponding author.

\section*{Additional information}
All data, models, generated or used during the study appear in the submitted article, code generated during the study are available from the corresponding author by request.

\section*{Supplementary information}
\subsection{The 8-Lorenz system}
\label{sec:app1}
\begin{figure}[htbp]
	\centering
	\includegraphics[height=8cm,width=0.7\textwidth]{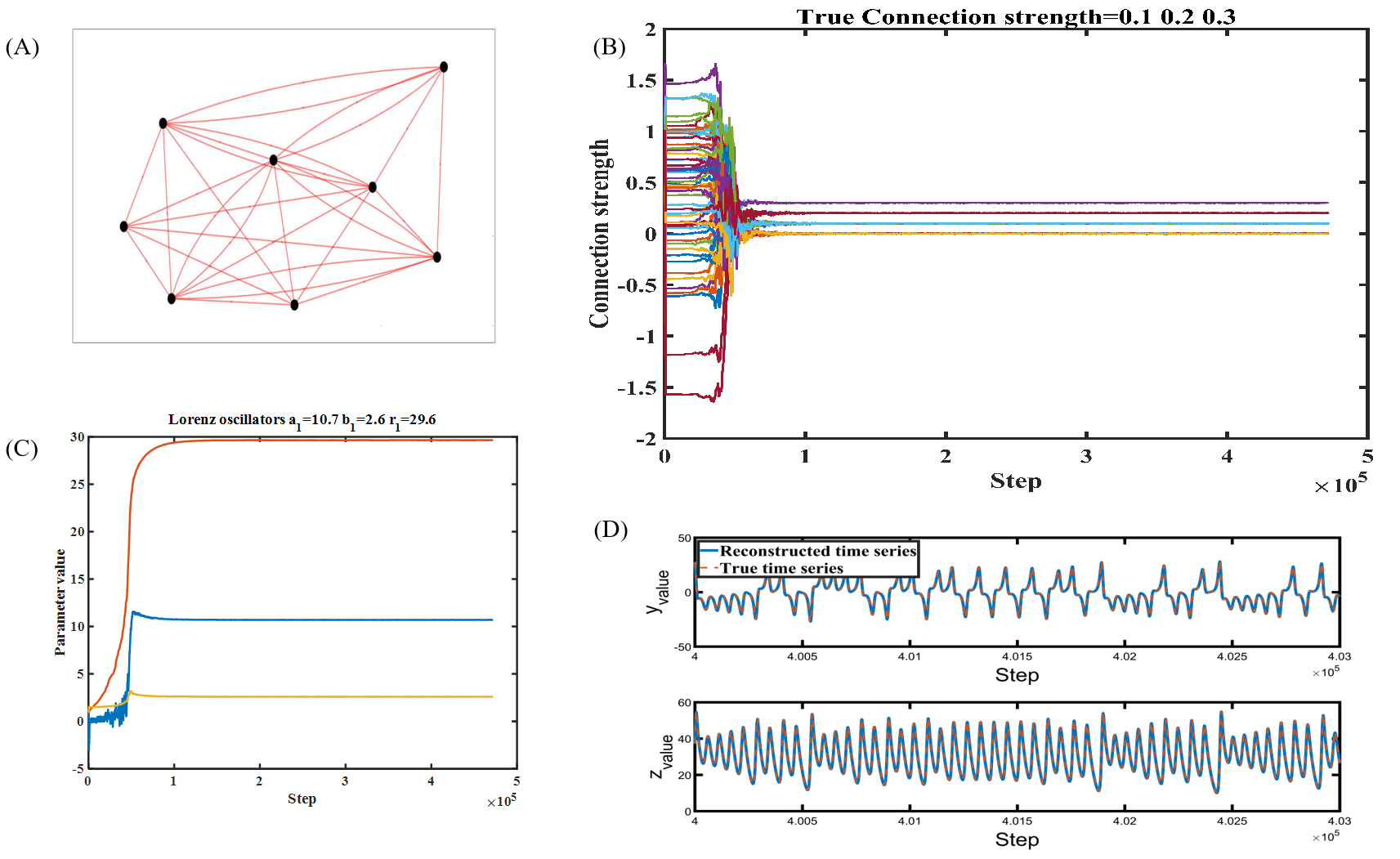}
	\caption{Reconstruction in a network of $8$ coupled Lorenz systems with different local parameters. (A) The coupling network. (B) Inference of the coupling strengths. (C) Local parameter reconstruction of one node.(D) Reconstruction of the time series of the hidden variable y and z of that node.}
	\label{fig:lorenn=8}
\end{figure}

Here we choose an \text { Erdős-Rényi (ER)} network of 8 nodes as an example, depicted in Fig.~\ref{fig:lorenn=8}(A). The coupling strengths between nodes are randomly selected from $\{0.1, 0.2, 0.3\}$.
For small networks with unknown structures, we may assume full connection and identify true links based on the construction result. The inference of coupling strengths with our method is shown in Fig.~\ref{fig:lorenn=8}(B). It can be seen from the figure that with the evolution of inference equations, the coupling strengths change greatly from the initial value $1$ at the very start, and converge to the correct values after the transient. When searching in a huge parameter space, the parameters are wandering a lot at the beginning and the error function $\Delta$ is not always decreasing (Supplementary Fig.~\ref{fig:error}) due to the state change over time. Nevertheless, once the parameters enter the basin of attraction, it will be quickly attracted to the correct values which minimize $\Delta$. Very small inferred strengths  indicate no connection between the corresponding nodes. In Fig.~\ref{fig:lorenn=8}(C,D) the parameters of one node and the reconstructed time series of the variable $y$ and $z$ are displayed. Just like a single Lorenz system, the inferred results are quite accurate. 

\subsection{Recycling of data}
\label{sec:app2}
\begin{figure}[!htbp]
	\centering
	\includegraphics[height=8cm,width=0.7\textwidth]{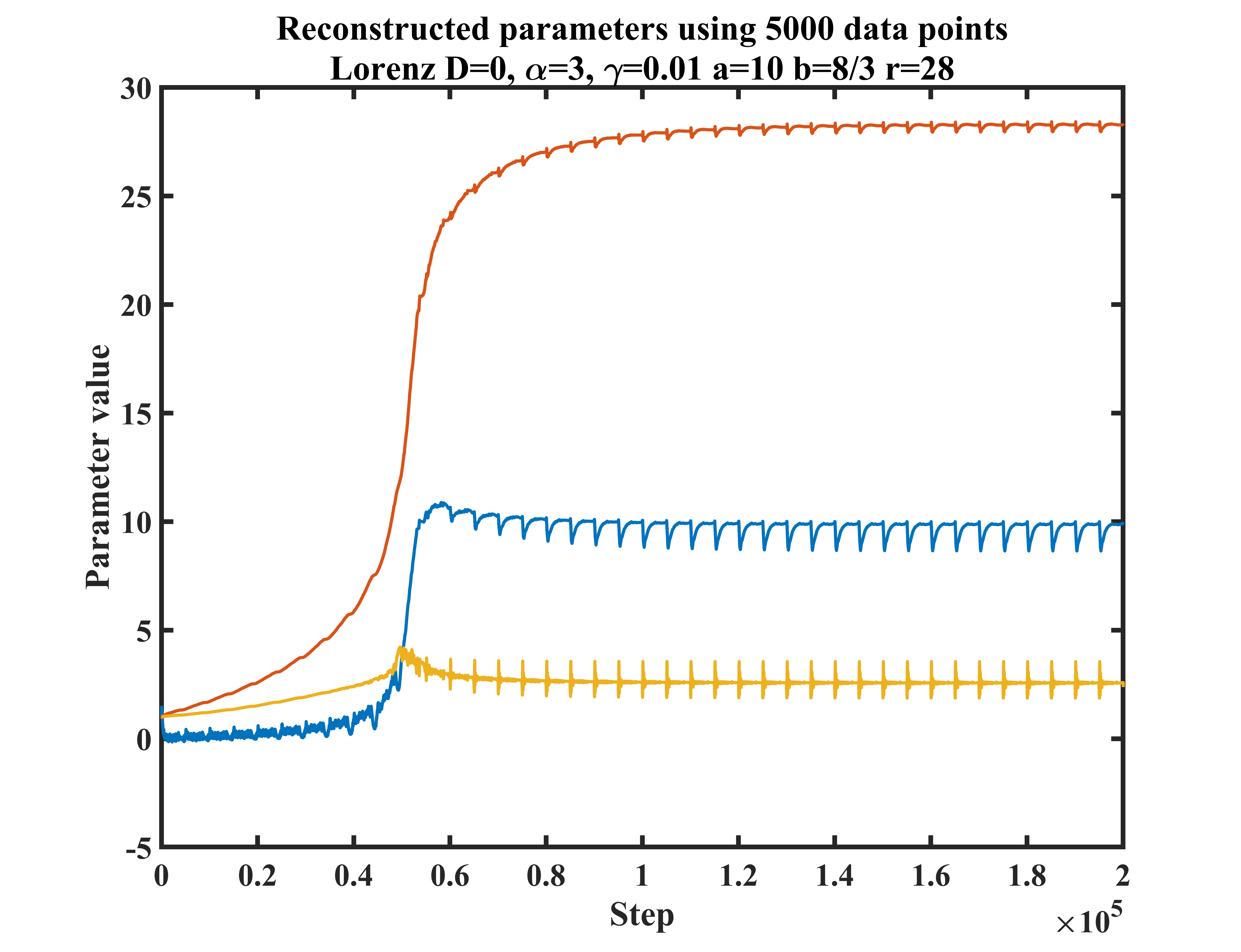}
	\caption{Reconstruct the Lorentz system by recycling $5000$ data points.}
	\label{fig:para_5000data.png}
\end{figure}

In our reconstruction process, the amount of data needed is very large. For example, for the Lorentz system we need $50,000$ sampling points to recover the dynamics. What if there are not so many sampling points? Our solution is to recycle the measured data to generate longer orbits. As shown in Fig.~\ref{fig:para_5000data.png}, we only used $5000$ data points for the reconstruction. It seems that all the unknown parameters converge to the right values but with jittering appearing every $5000$ points. This is because the recycled orbit breaks every $5000$ points and so the continuous evolution is restarted with a jump at this point. Nevertheless, the quick relaxation of parameter values before next jump is clearly seen, which indicates the robustness of the reconstruction and feasibility of recycling the data. 
\subsection{Error function $\Delta$ in Eq.~\ref{eq:f2}}
\begin{figure}[htbp]
	\centering
	\includegraphics[height=8cm,width=0.7\textwidth]{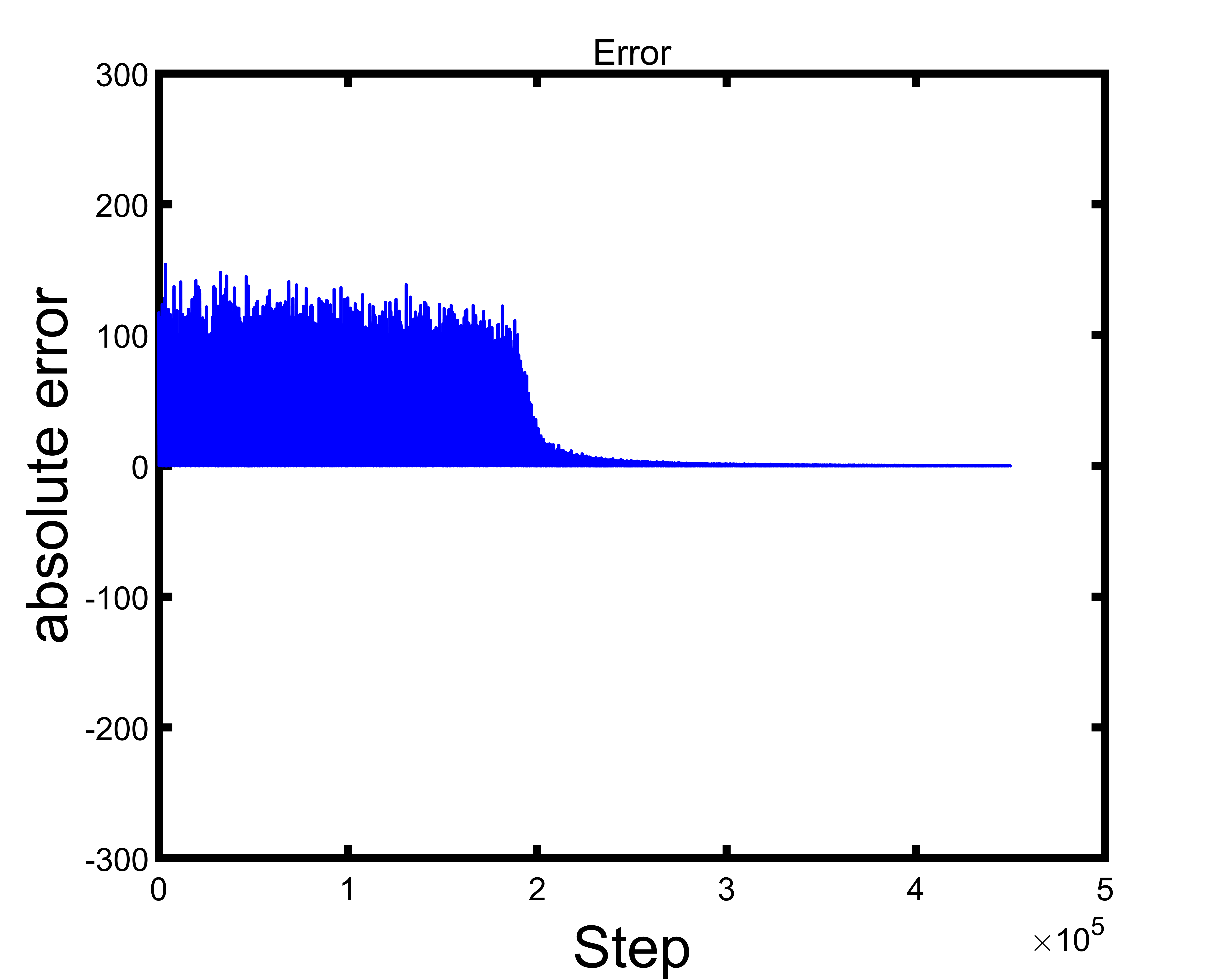}
	\caption{The error function $\Delta$ in the 8-Lorenz system.}
	\label{fig:error}
\end{figure}

Although the inference system evolves along the negative gradient, the error function $\Delta$ defined in Eq.~\ref{eq:f2} does not always decrease since it is explicitly a time-dependent function. As shown in Fig.~\ref{fig:error}, at the initial stage, the error $\Delta$ indeed fluctuates a lot. But after entering the basin of attraction of the parameters and the dynamics, the convergence dominates the time evolution so that the error suddenly drops to almost $0$ as seen in the Figure. 

\subsection{Gene regulation dynamics}
\begin{figure}[htbp]
	\centering
	\includegraphics[height=5cm,width=0.7\textwidth]{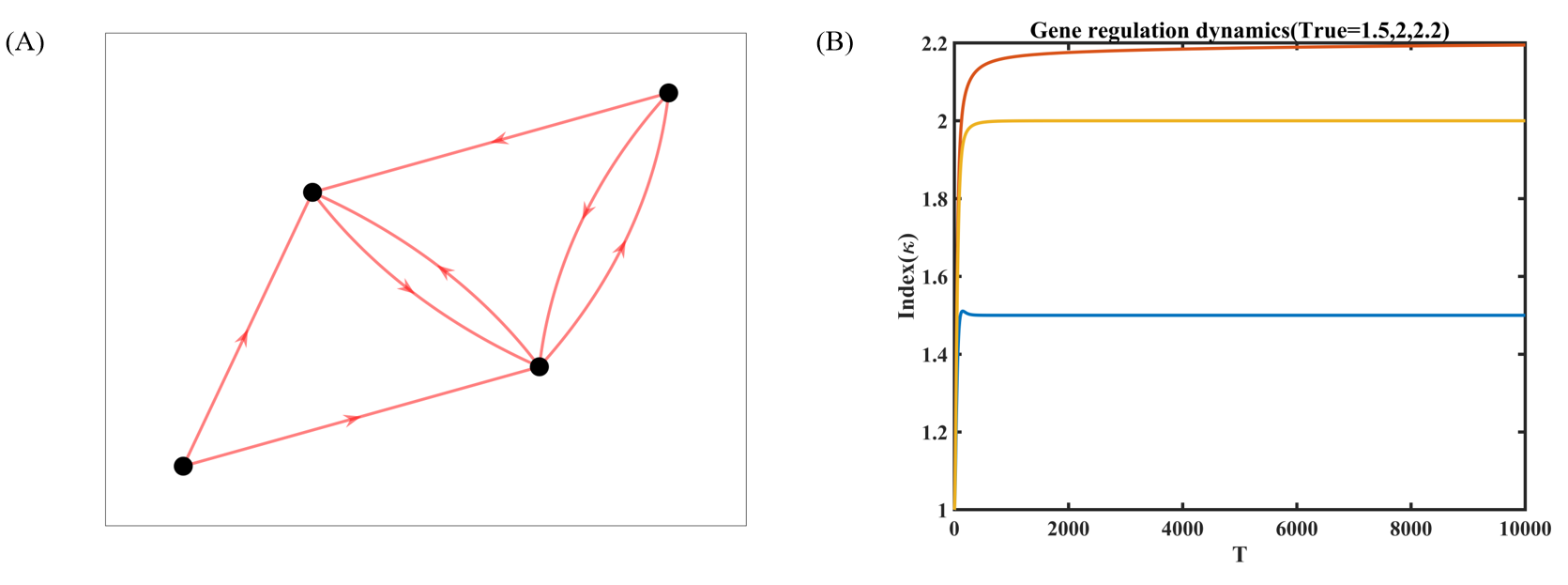}
	\caption{Reconstruction in gene regulation networks. (A) A gene regulation network.(B) Inference of Hill's Coefficients. }
	\label{fig:gene}
\end{figure}

Due to its evolution character, the current algorithm may treat dynamics with a nonlinear parameter dependence. To demonstrate this salient point, we use a simple gene regulatory network~\cite{2009Reconstructinggene}.
\begin{equation}
\frac{d g_{i}}{d t}=a_{i} g_{i}+\sum_{j=1}^{N} A_{i, j} \frac{g_{j}^{\kappa_j}}{g_{j}^{\kappa_j}+1}
\label{eq:gene}
\end{equation}
where $g_{i}$ is the concentration of gene $i$ , $\kappa_i$ denotes the Hill's coefficients, and $a_{i}\in[-0.3,-0.1]$. Parameters $A_{i, j}$ denotes the regulation strength of gene $j$ on gene $i$ which is set to $0.1$ and the network is plotted in Fig.~\ref{fig:gene}(A). Given the time series of $g_i$, we would like to deduce the connection $A_{ij}$ and the Hill's coefficients in the equation, which take three different values. To start the reconstruction, all the values of the parameters are initially set to $1$ in the inference equation. The inference dynamics is depicted in Fig.~\ref{fig:gene}(B).  It is easy to see that the unknown parameters are quickly driven to the true values, although they appear as power indicies in Eq.~(\ref{eq:gene}). Hence, equipped with the current reconstruction technique, it seems possible to tackle intricate dynamics with nontrivial parameter dependence in real-world applications.

\subsection{Lyaponov exponent {\em vs} the decay rate $\alpha$}
\begin{figure}[htbp]
	\centering
	\includegraphics[height=5.7cm,width=0.8\textwidth]{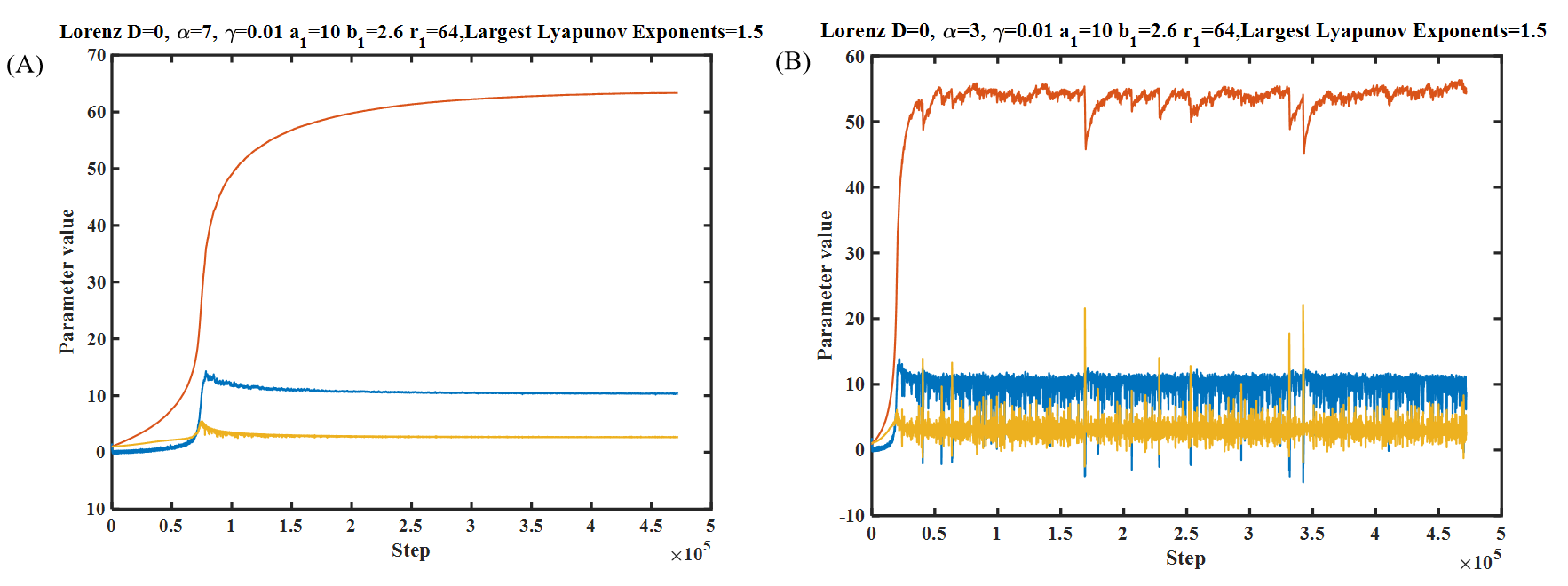}
	\caption{A proper decay rate $\alpha$ depends on the Lyapunov exponent. Here $t_ {unit} = 0.01,\gamma=0.01$, and the noise strength $D = 0$. The largest Lyapunov exponent is $1.5$. (A) The inference at $\alpha=7$, (B) the inference at $\alpha=3$.}
	\label{fig:alpha7}
\end{figure}

The choice of the decay rate $\alpha$ is closely related to the Lyapunov exponent. To compare Fig.~\ref{fig:f1} and \ref{fig:f2}, we increase the parameter $r$ in the Lorentz system~(\ref{eq:lorenz}) from $28$ to $64$, and as a result the largest Lyapunov exponent increases from $0.8$ to $1.5$. If we still take the value $3$ of the decay rate $\alpha$ as in Fig.~\ref{fig:f1}, the inference fluctuates a lot with no perceivable convergence as depicted in Fig~\ref{fig:alpha7} (B). In fact, to ensure the convergence, a larger $\alpha=7$ is needed to counter the increased instability as shown in Fig~\ref{fig:alpha7} (A). 

\begin{figure}[!htbp]
	\centering
	\includegraphics[height=9cm,width=0.8\textwidth]{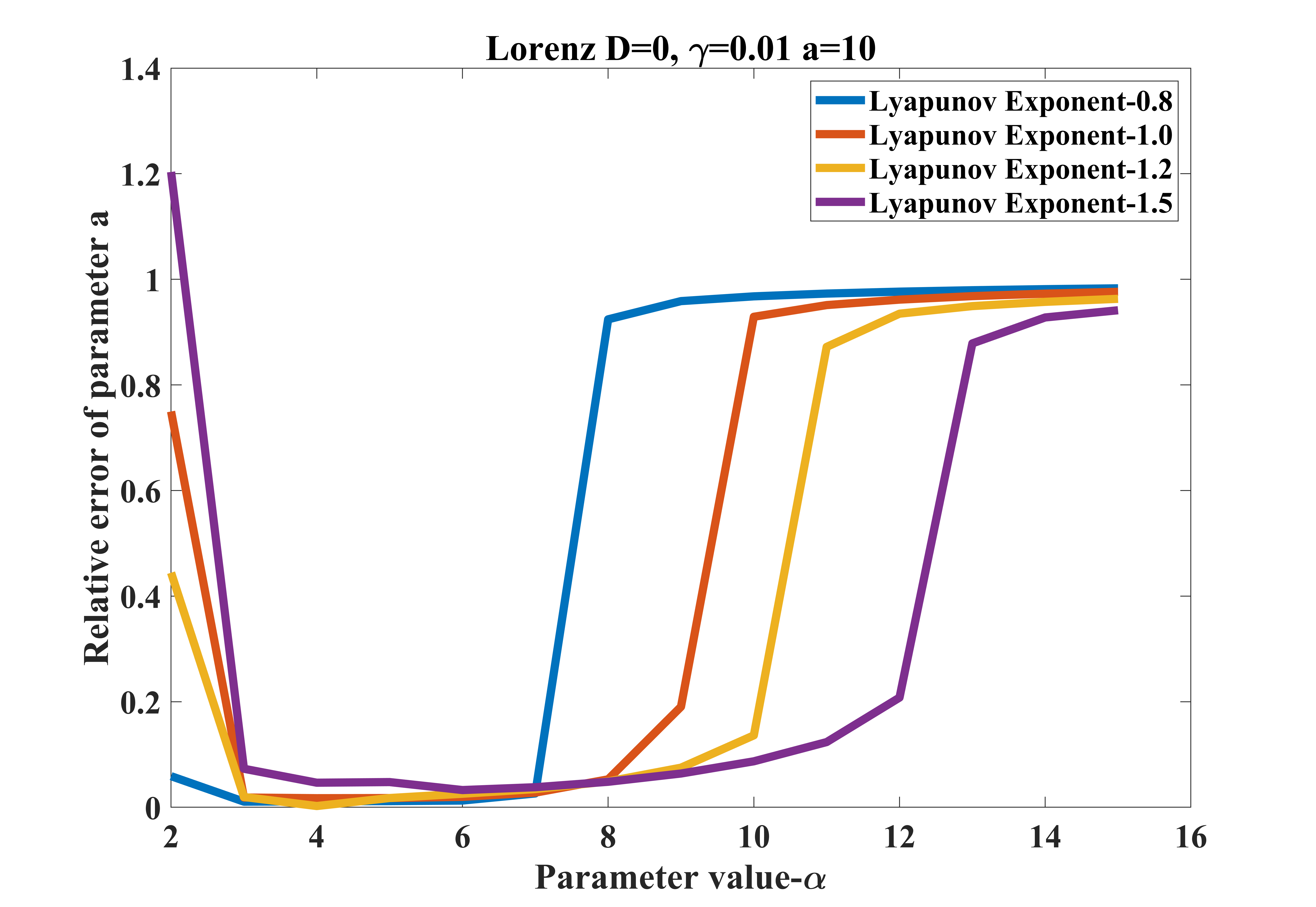}
	\caption{The errors in parameter reconstruction in the Lorenz system at different decay rates $\alpha$ for different Lyapunov exponents.}
	\label{fig:errorlya}
\end{figure}

In Fig.~\ref{fig:errorlya}, a more comprehensive comparison is made at different decay rates for different Lyapunov exponents. The effective $\alpha$ tends to increase with the Lyapunov exponent, being consistent with the above observation. It is good to note that the increase is slow.  There seems exist an interval such that the $\alpha$ in this interval is generally good for different $\alpha$ values.

\subsection{FHN Synchronized dynamics}
\begin{figure}[htbp]
	\centering
	\includegraphics[height=9cm,width=0.8\textwidth]{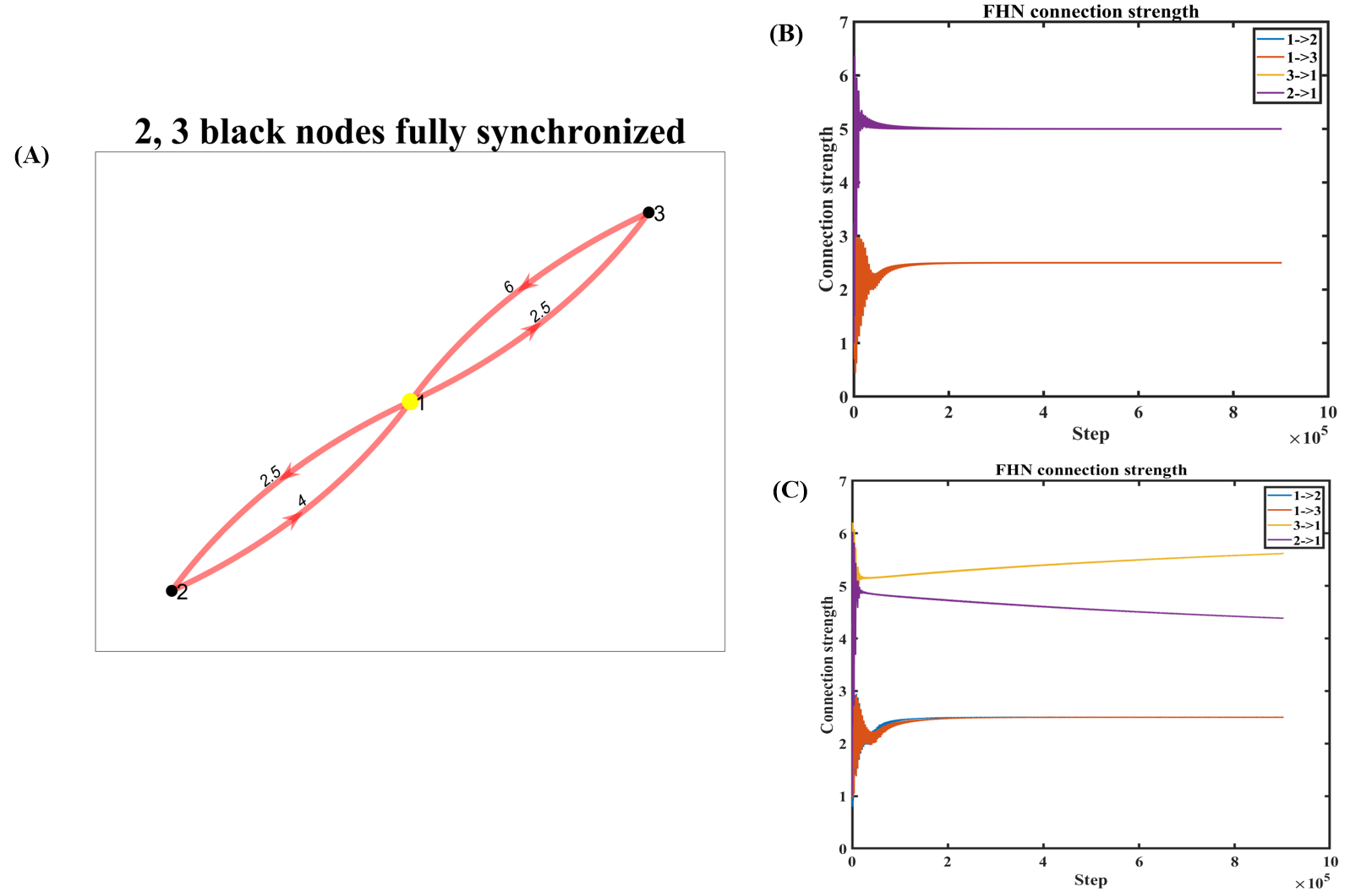}
	\caption{Reconstruction against synchronization in a network with three FHN nodes. (A) The network with the coupling strength=[2.5 2.5 4 6], $a_{i}=0.3,b_{i}=0.5,c_{i}=-0.04$.  Inference of the coupling strength in full synchronization (B) and in partial synchronization (C), with $a_{1}=0.3,a_{2}=0.4,b_{1}=0.5,b_{2}=0.5,c_{1}=-0.04,c_{2}=-0.03$.}
	\label{fig:FHNsyn}
\end{figure}

Synchronization generally makes the inference harder, which can be clearly seen here from a simple example. As shown in Fig.~\ref{fig:FHNsyn}(A), the synchronization dynamics of the three FHN nodes are used in the reconstruction. When node 2 and 3 are fully synchronized, it is impossible to distinguish them only from the time series of node 1 (Fig.~\ref{fig:FHNsyn}(B) ). Therefore, in this case, the obtained coupling strength $5$ is the average of the original two: $4$ and $6$. If they are not fully synchronized (by changing the parameter values), our method works but is slow to infer the correct coupling strength when close to synchronization (Fig.~\ref{fig:FHNsyn}(C).

\end{document}